# Electronic noise for Quantum Electronics

**J. I. Izpura**

**Abstract**

In 1951 it was proposed that the noise of a system would be due to the incessant dissipation of its thermal fluctuations of energy. Therefore, Johnson noise should reflect an incessant dissipation of fluctuations of electrical energy in resistors. The (cause-effect) model that we proposed for the capacitive energy held by their two terminals, keeps this fluctuation-dissipation dynamic. In this model, thermal translocations of single electrons between their terminals are the instantaneous fluctuations of energy that are dissipated incessantly by the collective reaction of all the carriers of this device. From this model whose displacement currents are noisy, we show that conduction ones must be noiseless, as it occurs in biased and unbiased resistors in thermal equilibrium.

## 1- Introduction

In optical oscillators (i. e. lasers), stimulated emission sustains oscillations while spontaneous emission is thermal noise that reduces its spectral purity. This broadening of their linewidth, whose counterpart is phase noise in electronic oscillators [1], makes very interesting our model for electronic noise based on spontaneous translocations of single electrons between terminals of resistors [2, 3]. Although capacitance shunting resistors is quite unknown regarding noise [4] its retarding effect on voltage signals trying to appear between its terminals caught my attention at an early age. I mean the capacitance $C_0$ that exists between any two terminals at distance $d$ in vacuum, or $C_{mat}>C_0$ when conductive material exists between them. Since this capacitance will shunt any two-terminal device (2TD) we can make, I kept it in mind revisiting Johnson noise, which should come from a "*thermal agitation of electric charge*", following the titles of [5, 6].

Since $C_{mat}$ holds the noise voltage of each resistor in thermal equilibrium (TE) at temperature T, I asked myself if its Johnson noise $v(t)$ could be the effect of fluctuating energy in $C_{mat}$. In this case, the density 4kTR $V^2$/Hz of a resistor of resistance R (k is the Boltzmann constant) would not be flat up to $f_Q$=kT/h [6], the limit of its $S_I$=4kT/R $A^2$/Hz density of Nyquist noise. Filtering $S_I$, the admittance $Y_{RC}(j\omega)$ of each resistor would give its Lorentzian spectrum of Johnson noise of amplitude 4kTR $V^2$/Hz and cut-off frequency $f_c=1/(2\pi RC_{mat})$ Hz. Integrating over all frequencies this density, the mean square voltage of Johnson noise is: $<v^2(t)>=kT/C_{mat}$ $V^2$, as thermal equipartition (TEQ) demands in $C_{mat}$. The model for the electronic noise of resistors described in [2, 3] was proposed from these ideas and the impulse response $h(t)$ associated to $Y_{RC}(j\omega)$.

Taking electrons as individual quanta of charge, $C_{mat}$ allows sudden translocations of single electrons between terminals of a resistor at the cost of tiny packets of energy $E_1=q^2/(2C_{mat})$ taken from its thermal bath [3]. Thus, $C_{mat}$ not only allows this interaction but also quantizes its energy $E_1$. This is the fluctuation of energy that enters the resistor causing a voltage shift $\Delta V=\pm q/C_{mat}$ volts and null risetime that decays with time constant $\tau=RC_{mat}$ as $E_1$ is dissipated. The decay $h(t)=(q/C_{mat})\times\exp(-t/\tau)$ volts is the response of the resistor to this fluctuation. This is the effect (reaction) caused by each fluctuation called "thermal action" before meeting [7] during my interaction with authors of [4] in 2008. Integrating in time the power $|h(t)|^2/R$, one gets the energy $E_1$ set in $C_{mat}$ that is dissipated subsequently by the current $i(t)=h(t)/R$. Johnson noise is thus a random voltage coming from the $h(t)$ responses to fluctuations that occur in resistors at random instants and with random signs. This summarizes our model for electronic noise agreeing with [7].

From $<v^2(t)>=kT/C_{mat}$, a resistor in TE dissipates $P_{TE}=kT/\tau$ watts on average by its Johnson noise driving its resistance R. Since this occurs under open circuit conditions, $P_{TE}$ must come from electrical power entering $C_{mat}$ as fluctuations of $E_1$ joules each, that occur with mean rate $\lambda$. To sustain $P_{TE}$ as their mean effect, the rate $\lambda$ must be [2, 3]:

$$P_{TE}=\frac{kT}{RC_{mat}}=\frac{\lambda\times q^2}{2C_{mat}}\Rightarrow\lambda=\frac{2kT}{q^2R}=\frac{2kT}{q^2}G\Rightarrow G=\lambda\frac{q^2}{2kT}\qquad(1)$$

We envisaged Eq. (1) at low T ($T\rightarrow 0$) to give room for the dissipation of each $E_1$ between fluctuations. At higher T's, however, this removal of $E_1$ also occurs if we use material of low relaxation time $\tau_d$ to have a low $\tau=RC_{mat}$ for this device. In both cases, the rate $\lambda$ does not depend on $C_{mat}$ and $G=1/R$ (the readiness of the resistor to dissipate energy) is proportional to $\lambda$. Doubling $E_1$ by halving $C_{mat}$ does not affect $\lambda$, but doubles the mean square voltage noise: $<v^2(t)>=kT/C_{mat}=4kTR\times BW_n$ of the low-pass, R-C filter that each resistor is, since we double its noise bandwidth: $BW_n=(\pi/2)\times f_c$ Hz. In addition to the rate of fluctuations in the resistor, Eq. (1) suggests that its conductance G reflects its ability to dissipate energy $\lambda$ times per second too. This view on G helped to propose a model for conduction current devoid of charge displacements other than those giving rise to Johnson noise (see the Annex). The Johnson noise of a 1MΩ resistor at room T would come from its $\lambda\approx 3{,}2\times 10^{11}$ fluctuations per second, looking like those of Fig. 10 of [2], whose energy $E_1$ is well dissipated before the next fluctuation occurs. Fig. 9 of [2] shows the Lorentzian spectrum of each response $h(t)$, thus the spectrum of the Johnson noise they produce.

For people used to Johnson noise of flat spectrum I will say that its $BW_n>100$ GHz is more the rule than the exception for conductive materials (see below). This ends the introduction to our model for electronic noise, where the minute charge $q=-1{,}6\times 10^{-19}$ C of an electron allows fluctuations needing tiny amounts of thermal energy to occur in capacitive paths. For $C_{stray}\approx 0{,}5$ pF shunting a resistor at room T $E_1=(q^2/2)/(C_{mat}+C_{stray})$ gives $E_1\approx 0{,}16\mu eV$, which is six parts per million $kT\approx 26meV$. This $E_1<<kT$ agrees with the huge rate $\lambda$ of Eq. (1) for resistors made from suitable material. Since this suitability comes from its relaxation time $\tau_d$, I will consider this subject in Section 3 after the issues about shot noise of Section 2. The likelihood of shot noise due to conduction current is dealt with in Section 4 to pave the way for Section 5, where the assignment of shot noise to this dissipative current is considered from our model for electronic noise.

## 2- Shot noise: present in capacitors and elusive in resistors?

Looking for works on shot noise around 2008, I confirmed that this noise, studied in 1918 by W. Schottky, is assigned to discrete charges like electrons of charge $-q$ C that pass between two terminals. For the average passage of $n$ electrons per second, $S_{Ish}=2q^2n$ A$^2$/Hz is the unilateral density of shot noise at low frequency (i. e. as $f\rightarrow 0$). If the current $I_C$ biasing a resistor was the mean passage of $n=I_C/q$ electrons/second, its density of noise voltage between its terminals should be: $S_{Vtot}=(4kTR+2qI_CR^2)$ V$^2$/Hz because at low $f$, its resistance R *converts* $S_{Ish}$ into density of noise voltage. Amazingly, the noise found in this device thus biased only is: $S_V=4kTR$ V$^2$/Hz, which is the Johnson noise measured in TE for $I_C=0$. And most people are unaware of this lack of shot noise assigned to bias currents, as I was in the eighties. This work gives a good reason for this fact.

Textbooks taking a current $I_C$ as a flow of discrete electrons give $S_{Ish}=2qI_C$ A$^2$/Hz as its density of shot noise as $f\rightarrow 0$. Assuming this flow we accept $S_{Ish}$ and its notion of charges in transit. From this idea (that works when electrons cross vacuum regions or space-charge ones in solid-state devices) the shot noise of $I_C$ should be: $S_{Ish}=2qI_C$ A$^2$/Hz. But to contend that this is true, we must find $S_{Ish}$ in experiments. This requires converting $S_{Ish}$ into a voltage density $S_{Vsh}$ (its effect) that, once measured by a voltmeter, allows to infer the current of density $S_{Ish}$ that has been its cause. To know a current, this conversion and inference both are required.

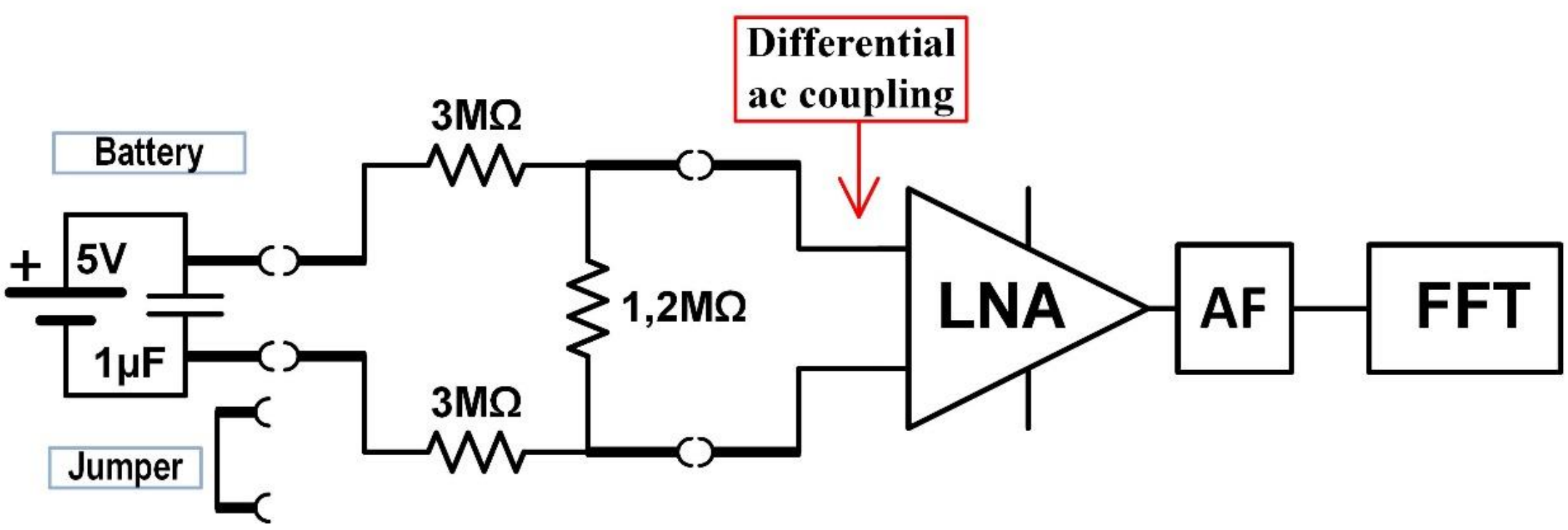

**Fig. 1**. Low-Noise Amplifier EGG-PAR-113 driven to show that the presence of a bias current $I_C\approx 0{,}7\mu$A in a 1MΩ resistor at room T, does not increase its Johnson noise. Note that $R_t$=1MΩ shunted by $C_{stray}$ of the differential input of this LNA adds a third low-pass filter to the two antialiasing ones (AF) of $f_c$=10kHz that uses our FFT analyzer sampling at 250 ksamples/second (see the text).

This conversion of $S_{Ish}=2qI_C$ A$^2$/Hz into $S_{Vsh}=2qI_CR^2$ V$^2$/Hz as $f\rightarrow 0$ in a resistor is granted by its own resistance R. Thus, let me use the circuit of Fig. 1 to show the lack of shot noise assigned to a current $I_C\approx 0{,}7\mu$A biasing its resistor of $R_t\approx$1MΩ (1,2MΩ shunted by 6MΩ). A 5V battery shunted by 1μF (that grants its low enough impedance at $f$=10Hz) allows to set $I_C$=5V/7,2MΩ in the three resistors of Fig. 1. Note the symmetry of this circuit to exploit the differential input of our EGG-PAR-113, low noise amplifier (LNA). By a small jumper (i. e. a short) connecting the two 3MΩ resistors we will measure the Johnson noise of $R_t\approx$1MΩ. Replacing this jumper by the battery, the dc current in the two resistors of 3MΩ in series becomes: $I_C$=5/7,2=0,7μA The Nyquist density of this 6MΩ resistor is: $S_{I2}=4kT/R_t=2{,}8\times 10^{-27}$ A$^2$/Hz whereas the shot noise "expected from its dc $I_C$" (better said: "assigned to its dc $I_C$") is: $S_{Ish2}=2qI_C=2{,}25\times 10^{-25}$ A$^2$/Hz, thus: $S_{Ish2}\approx 80S_{I2}$.

Concerning the resistor of R=1,2MΩ that shunts those 6MΩ, its Nyquist density $S_{I1}=4kT/R=1{,}4\times 10^{-26}$ A$^2$/Hz is 16 times lower than $S_{Ish1}=2qI_C=2{,}25\times 10^{-25}$ A$^2$/Hz (the shot noise assigned to its $I_C$) thus $S_{Ish1}=16S_{I1}$. The total noise current density between terminals of Fig. 1 will be: i) $S_I=4kT/R_t=1{,}66\times 10^{-26}$ A$^2$/Hz (Nyquist density of $R_t$) plus ii) the shot noise assigned both to the dc $I_C$ on 6MΩ: $S_{Ish2}=2qI_C=2{,}25\times 10^{-25}$ A$^2$/Hz and that assigned

to $I_C$ on the 1,2MΩ resistor that also is: $S_{Ish1}$=2$q$$I_C$=2,25×$10^{-25}$ $A^2$/Hz. The total density of shot noise *assigned to $I_C$* is: $S_{Ish}$=2×2$q$$I_C$=4,5×$10^{-25}$ $A^2$/Hz. This gives $S_{Ish}$=27$S_I$, and since ($S_I$ + $S_{Ish}$)=28$S_I$ entails 14,5 dB more noise power for $I_C$=0.27µA than for $I_C$=0, this shot noise assigned to $I_C$ should be much greater than that for $I_C$=0 in TE. The point, however, is that our measurements of Fig. 2 confirm unambiguously and one more time, that *we should not assign shot noise to a bias current* like $I_C$.

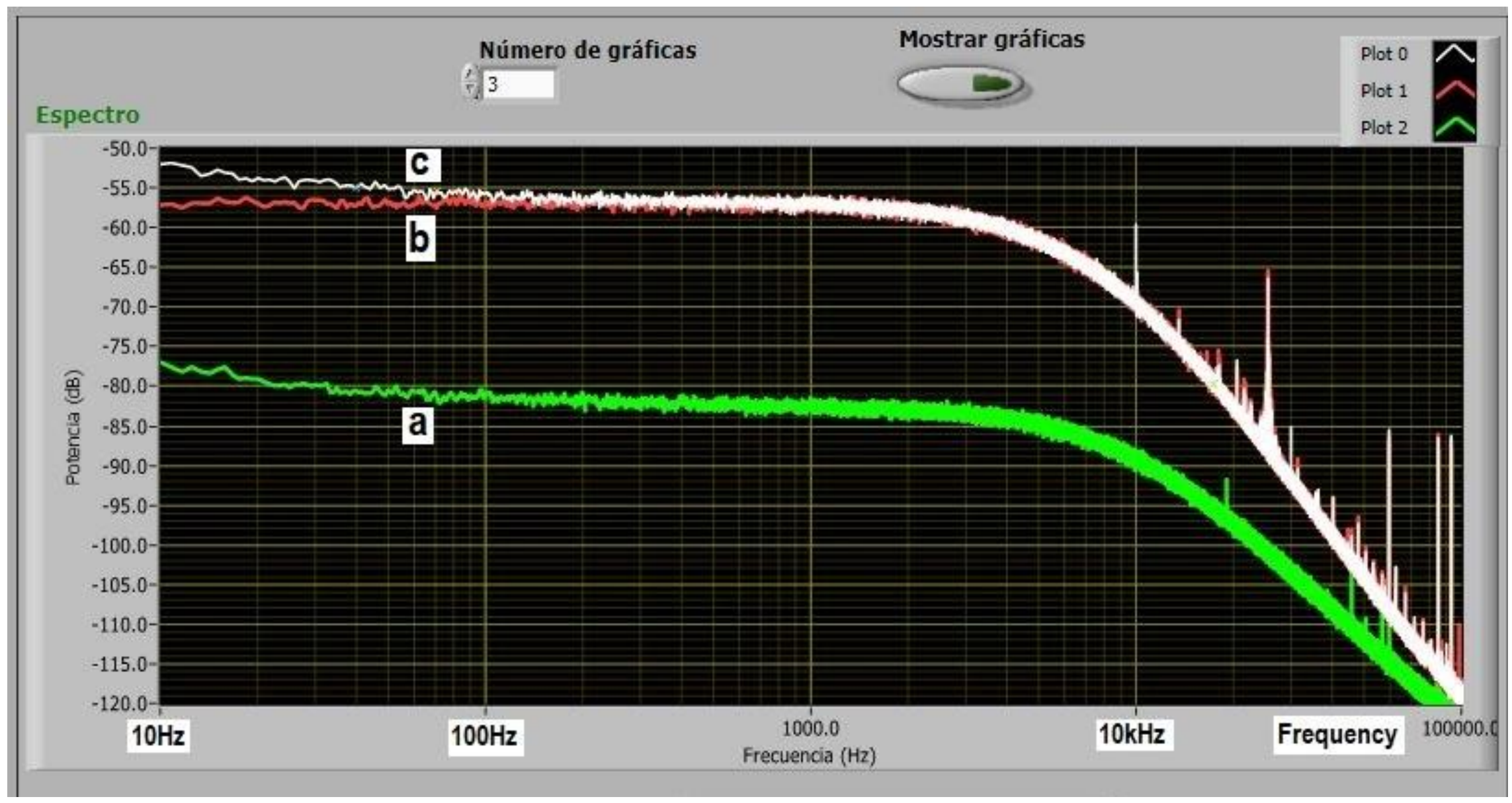

**Fig. 2**. Noise voltage measured in the circuit of Fig. 1 in these cases: a) noise floor with the inputs of the LNA shorted together in Fig 1; b) Noise voltage of the $R_t$=1MΩ resistor appearing at the input of this LNA when the jumper is connected ($I_C$=0); c) Noise voltage at the input of this LNA when the battery is connected instead the jumper ($I_C$≈0,7µA).

Concerning the measurements of Figs. 2 and 3, they are densities of noise voltage in $V^2$/Hz that do not consider the voltage gain ($10^4$ times, 80dB) set in our LNA. As we will show, the direct measurement of densities of noise current is unreal. We must convert them into voltage densities that we can measure directly by a voltmeter like our spectrum analyzer. Although I have made calculations with densities of noise current in $A^2$/Hz, this is due to the nature of the problem with all these densities adding in parallel. What we see in Figs. 2 and 3 is voltage density at the input of the LNA. Going to Fig. 2, its red graph b) is the voltage noise density of the $R_t$=1MΩ resistor of Fig. 1 with the jumper connected. This is the Johnson noise of this $R_t$=1MΩ resistor that would be: $S_V$=4kT$R_t$=1,66×$10^{-14}$ $V^2$/Hz (-137,8 dB). Considering the 80dB gain of our LNA, the flat region of curve b) at -57dB goes to (-57-80)=-137dB. Although adjusting the continuous gain of this LNA, I could set exactly this flat region at -137,8 dB, I prefer to use its fixed gain of 80dB because the overwhelming proof that I am giving does not need such calibration accuracy.

Going back to Fig. 2, its curve b) in red is the density of voltage noise $S_V$ of the resistor of Fig. 1 with $I_C$=0, thus its Johnson noise in TE at T=300K. By contrast, its curve c) (white) is the density of voltage noise of this resistor of Fig. 1 with $I_C$=0,7µA when the battery replaces the jumper. The proof I refer to is that graphs b) and c) are equal between 100Hz and 10kHz, thus indicating that $I_C$=0,7µA does not change the Johnson noise of our 1MΩ resistor of Fig. 1. With this resistor biased by $I_C$≈0,7µA, its density of noise voltage is far away from $S_{Vtot}=S_V+27S_V$ that one expects from assigning shot noise to its $I_C$. The $S_{Vtot}\approx S_V$ that we have found means that the bias current $I_C$ does not affect the "*agitation of charge*" [5, 6] giving rise to its Johnson noise. By contrast, $S_{Vsh}\neq 0$ is found easily in devices like junction diodes where displacement currents take place [2]. Thus, displacement currents give shot noise that conduction currents do not give and because resistors offering pure resistance are unreal, I will use their admittance $Y_{RC}(j\omega)$ that gives room for displacement and conduction currents between their terminals.

Before using $Y_{RC}(j\omega)$, let me say that graph a) of Fig. 2 is the noise of the LNA with its inputs short to ground. To make this paper short I will say that the low value of graph a) (green) respect to graphs b) and c) means that the contribution of the LNA to the measured noise of graphs b) and c) is negligible. Although this wonderful LNA has lost part of the brilliant features it had forty years ago, it remains valid to measure the Johnson noise of a 1MΩ resistor (quite high) and even to show the small *excess noise* due to $I_C$ that curve c) shows below 100Hz. This white graph shows excess noise over the floor that the Johnson noise of curve b) in red is for this measurement (resistance noise revealed by the injected current $I_C$). Hence, *beware excess noise if you inject current in a resistor* (see below). Finally, Fig. 3 shows curve a) of Fig. 2 (now in red) which is found using two cascaded antialiasing filters of $f_c$=10kHz, whereas the same measurement with only one of these filters to allow some aliasing is shown in white. The green arrow indicating negligible aliasing effects under $f$=2kHz means that aliasing effects in Fig. 2 below 10kHz can be discarded due to its third antialiasing filter (see later).

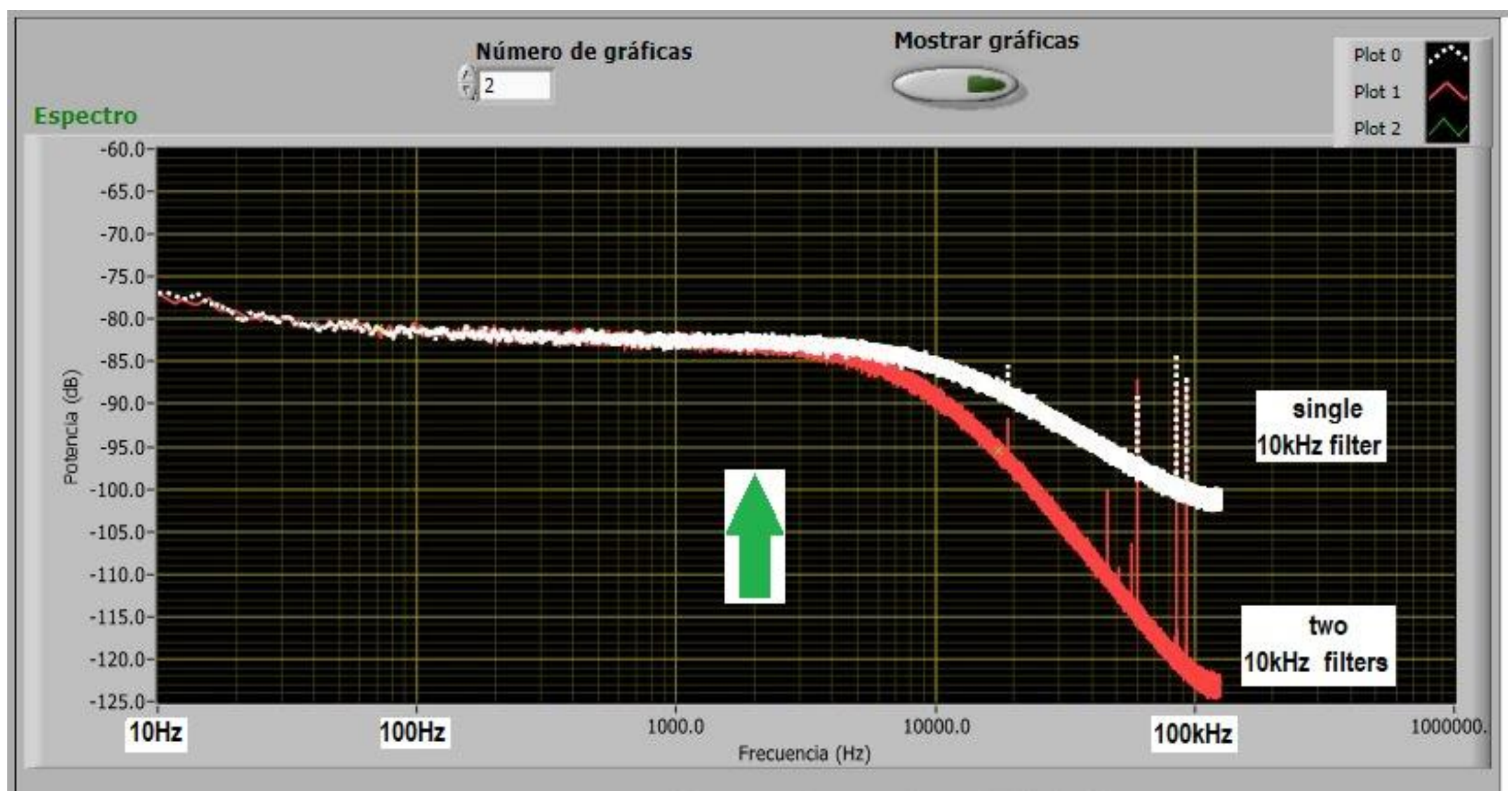

**Fig. 3**. Noise voltage measured in the circuit of Fig. 1 (noise floor) with the inputs of the LNA shorted together, using one and two antialiasing filters (AF) of $f_c$=10kHz (see the text). The third AF does not exist here because the null resistance of the shorted inputs of the LNA in this case.

Following [2, 3], Johnson noise would not exist without the $C_{mat}$ that holds the fluctuating energy that it reveals. To shed light on the noiseless nature of bias currents in resistors, let me show first where their Johnson noise comes from. Next, I will show that assigning shot noise to conduction currents is unfounded. Concerning the dissipation in a resistor, I will consider that it is the conversion of electrical energy into vibrational energy of its material called heat. I will also consider its noise voltage *v(t)* as formed by a set of sinusoidal voltages of different frequency $f=\omega/(2\pi)$ that $Y_{RC}(j\omega)$ links with a similar set of sinusoidal currents whose number doubles to consider current *in-phase* and current *in-quadrature* with each sinusoid of voltage. Calling *i-v pair* to a couple of sinusoids of voltage and current of equal frequency *f*, dissipation comes from *i-v* pairs whose terms of voltage and current are in-phase, with amplitudes (V and I) linked by a resistance R=V/I or conductance G=I/V. Due to their "in-phase condition", the ratio $v(t)/i(t)$ at each instant of time also is R (or G) and thus, $G\times(v(t))^2$ or $R\times(i(t))^2$ both give the active power that is the rate of energy losing its electrical form as it is dissipated.

Because fluctuations refer to electrical energy that the noisy device holds, I will also consider the reactive power coming from *i-v* pairs whose sinusoids $v(t)$ and $i(t)$ are in-quadrature (i. e. ±90º-off), with amplitudes linked by susceptance B=I/V or reactance X=V/I. Note the readiness of $Y_{RC}(j\omega)=G(\omega)+jB(\omega)$ to deal with noise coming from a fluctuation-dissipation dynamic where displacement and conduction currents play very different roles. However, works proposing that conduction current also should generate shot noise have been published. Reading one where the excess noise of a bias current $I_C$ was claimed to be shot noise due to $I_C$ [4], I decided to write this work on the impulsive current of thermal origin in resistors whose shot noise would cause their Johnson noise looking like a "continuous" voltage that conceals its discrete origin.

To end this section, I will say that the open circuit conditions of a resistor while its Johnson noise $v(t)$ is being generated and measured, *in no way mean absence of current in it*. Following [3] its noise voltage $v(t)$ comes from charge relaxing in its volume $V_Q$ by two currents cancelling one to each other at each instant of time. I refer to a conduction current $v(t)/R$ of equal magnitude but opposed sign to $i_d(t)=C\times(dv(t)/dt)$, a displacement current caused by previous fluctuations (see Fig. 4 of [3]). This $i_d(t)$ is due to those slow recoils displacing back, with time constant $\tau=RC_{mat}$, each electron displaced instantly by a fluctuation. To get familiarity with both types of current, next Section begins with the current-to-voltage conversion achieved by *i-v* converters and ends with electrical effects coming from the Maxwell relaxation time $\tau_d$ of the material of a resistor.

## 3- Electronic noise and Engineering

In 1928, J. B. Johnson measured a small voltage between terminals of different resistors in TE, showing its thermal origin [5]. This noise voltage of mean square value $kT/C_{mat}$ $V^2$ [2, 3] is generated by this "device", not by its "resistance" that appears in the formula $S_V(f)=4kTR$ $V^2/Hz$. To work in parallel mode, I will use the equivalent Nyquist noise density $S_I=4kT/R$ $A^2/Hz$ [6] that H. Nyquist inferred from [5]. Given the value of the frequency $f_Q$ where $S_I$ drops due to quantum reasons ($f_Q\approx6$ THz at room T), the density $S_I$ is taken as constant (i. e. "flat") in electronics, although this is not so for $S_V(f)$. The relationship $S_V(f)=R^2\times S_I$ that comes from neglecting in a resistor the effects of its shunting $C_{mat}$ as $f\rightarrow0$, in no way means that such capacitance does not exist.

With this capacitance in mind, I revisited Johnson noise looking for its origin as the thermal agitation of electric charge written in the titles of [5, 6]. I was looking for the way electrical charges like electrons agitate in resistors to generate their Johnson noise. Particularly, I was looking for the transducer generating this small voltage from their thermal bath, thus converting thermal radiation pervading the volume $V_Q$ of each resistor into its voltage of Johnson noise. Since this "photogeneration" has been described in [3], I only will say that what a resistor offers between its terminals is noise voltage $v(t)$ and noise current $v(t)/R_t$ to its total load $R_t$ that includes its own resistance R. Nevertheless, it will not deliver by its terminals the instantaneous currents (fluctuations) that gave rise to the noise voltage *v(t)* in its capacitance $C_{mat}$ between terminals [2, 3].

Let me reiterate that the value of a current is inferred from the voltage it causes in a 2TD acting as a current-to-voltage ($i \to v$) converter. Although electrical capacitance is the fastest one that I know, let me leave aside this notion for a while to consider the $i \to v$ converters (i. e. transimpedance amplifiers) used to convert a current into a proportional, measurable voltage. Fig. 4 shows a resistor feeding its Johnson noise to the input of an operational amplifier (OA) whose feedback by $R_F$ allows to convert a current leaving the resistor into a voltage $V_0$ appearing at its output. Note the capacitance in red that I will exclude from the circuit in black that I will use to describe the $i \to v$ conversion that most people expect today from this circuit. Due to its negative feedback, this OA will reduce the voltage at its input $v_\varepsilon=(v_{(+)}-v_{(-)})$ and due to its huge gain ($A_V \to \infty$), it will keep the voltage $v_{(-)}$ of node A very close to the voltage $v_{(+)}$ of node B because for finite values of $V_0$ the error voltage $v_\varepsilon$ tends to be null: $V_0/(A_V \to \infty)=v_\varepsilon \to 0$.

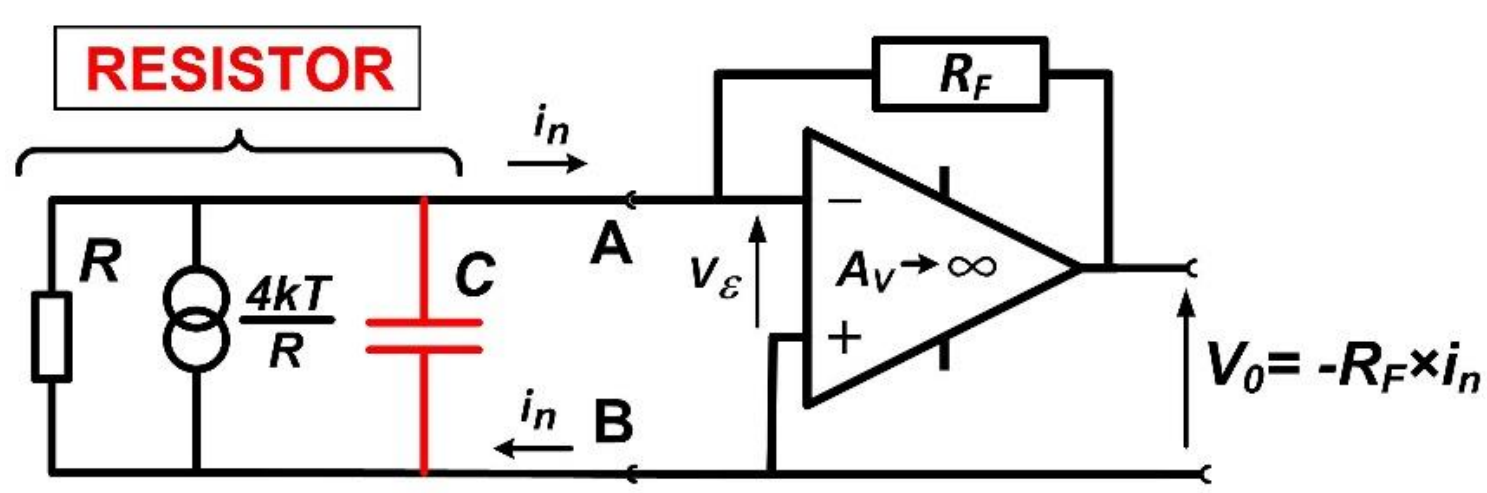

**Fig. 4**. Resistor feeding the input of an i-v converter that uses an Operational Amplifier with negative feedback by the resistor $R_F$. For a cogent modelling of the noise of the resistor, its capacitance in red should not be disdained (see the text).

A current $i_n$ arriving in node A from the resistor will try to raise its $v_{(-)}$ voltage. Reacting to this stimulus, the OA will give at its output a negative voltage $V_0=-(R_F \times i_n)$ to absorb such $i_n$ through its feedback resistor $R_F$. These $-(R_F \times i_n)$ volts at the output of the OA is the required voltage to keep null ($v_\varepsilon \to 0$) the voltage at the input of the OA. Since gain values $A_V>10^6$ V/V are usual in OAs like this one, the error voltage $v_\varepsilon$ in the circuit of Fig. 4, handling small signals like a Johnson noise at its input, will be under the μV. Thus, for currents like $i_n$, this is an *i-v* converter of conversion gain: $A_Z=V_0/i_n=-R_F$ volts per amp. For spectral densities that are proportional to the square of its output voltage $V_0$, the minus sign of this transimpedance $A_Z$ does not matter.

Years ago, I assumed that those currents generating $S_I$ in the resistor of Fig. 4 underwent the *i-v* conversion just described. Today, I have written: “to convert currents leaving the resistor into voltages…” because not every current in the resistor will leave it as the two currents $i_n$ of Fig. 4 suggest. Displacement ones giving rise to its $S_I$ in its $C_{mat}$, will not do it. Although a feedback $R_F$=10R will give in $V_0$ a density $S_V(f)$ ($V^2$/Hz) that numerically is: $-10 \times S_I=-40kT/R$ $A^2$/Hz, it is no more than the 4kTR $V^2$/Hz density that already exists between terminals of the resistor, inverted and amplified by ten. This occurs because $C_{mat}$ is a much faster $i \to v$ converter than the OA, which must wait until this voltage appears in $C_{mat}$ to react. This wait for the effect of each fluctuation makes useless any feedback circuitry to remove phase noise due to fluctuations in oscillators [2, 3, 8]. Thus, the series circuit of R and the density 4kTR $V^2$/Hz or the parallel density 4kT/R $A^2$/Hz shunting R, both need $C_{mat}$ to form a complete circuit of a resistor from the noise viewpoint.

The gain $A_Z=V_0/i_n=-R_F$ V/A in Fig. 4 results from assuming that $i_n$, the current leaving the resistor by its terminal A towards the (-) input of the OA, is deviated by the OA through $R_F$ to keep null its input voltage $v_\varepsilon$. This idea works for conduction currents that "obey voltages," but for displacement ones $C_{mat}$ makes a big difference. A first clue comes from taking as equipotential the A and B nodes connecting the noisy resistor to this *i-v* converter. Viewing the two $i_n$ currents of Fig. 4 as electrons moving in opposed senses along the wires A and B, we find a nonsense situation because equipotential wires forcing corpuscular electrons to move as the arrows labeled $i_n$ suggest sounds unlike. This led me to consider how an electron could pass between terminals of a 2TD while keeping current continuity and without infringing Special Relativity. The result is that it passes instantly by the capacitance of this 2TD, at the cost of a small thermal energy [3]. This suggested that the fluctuations of [2, 3] were displacement currents of thermal origin that never would leave the resistor of Fig. 4 as its $i_n$ arrows suggest.

Instantaneous displacement currents would not obey voltages like $V_0$ at the output of the OA "inviting them to pass across $R_F$" in Fig. 4. Each resistor offers a voltage $v(t)$ between terminals, thus delivering a conduction current $i(t)=v(t)/R_t$ to its total load $R_t$ (of many GHz if needed), but this $i(t)$ is totally different from those displacement currents that have generated $v(t)$ in its $C_{mat}$. As soon as a fluctuation updates $v(t)$ in $C_{mat}$ by a step $\Delta V=q/C_{mat}$ volts of null risetime, it already is gone. This is why we find noise voltage of $S_{Vt}(f)$ $V^2$/Hz as $f \rightarrow 0$, whose Lorentzian shape at high $f$ will be seen if our setup has enough bandwidth. Measuring $S_{Vt}(f)=4kTR_t$ we can infer $S_V(f)=4kTR$ $V^2$/Hz first, and next $S_I=4kT/R$ $A^2$/Hz as the density of noise current that shunting R, is equivalent to $S_V(f)$ in series with R. But taking Nyquist noise as a set of displacement currents that leave the resistor to generate a voltage of noise out of its shunting $C_{mat}$ is meaningless.

To have enough bandwidth in Fig. 4 let me replace its OA by a transmission line (TL) collecting the Johnson noise of its resistor to feed the input of a spectrum analyzer (SA) matched to its far end. Being lossless, this TL will have a real line impedance that will be: $R_0=\sqrt{(L/C_{TL})}$ where $L$ and $C_{TL}$ are its inductance and capacitance per unit length. The $dL$ elements of this TL of Fig. 5 suggest that the fluctuations of the resistor will not leave its $C_{mat}$ since they are not magnetic energy, thus leaving in $C_{mat}$ their responses $h(t)$ of $\pm\Delta V$ volts decaying in time as they excite its own R shunted by the $R_0$ of the TL. The random series of $h(t)$ pulses thus created make up the noise voltage $v(t)$ that injects into the TL the voltage and current waves of Johnson noise that go towards the resistor of the SA at its far end. Thus, the resistor feeds the TL with its noise $v(t)$ due to fluctuations in its $C_{mat}$ that decay incessantly with time constant $\tau=R_tC$, where $R_t$ is the resistance R of the resistor shunted by the line impedance of our TL that I will take $R_0=50\Omega$ hereafter.

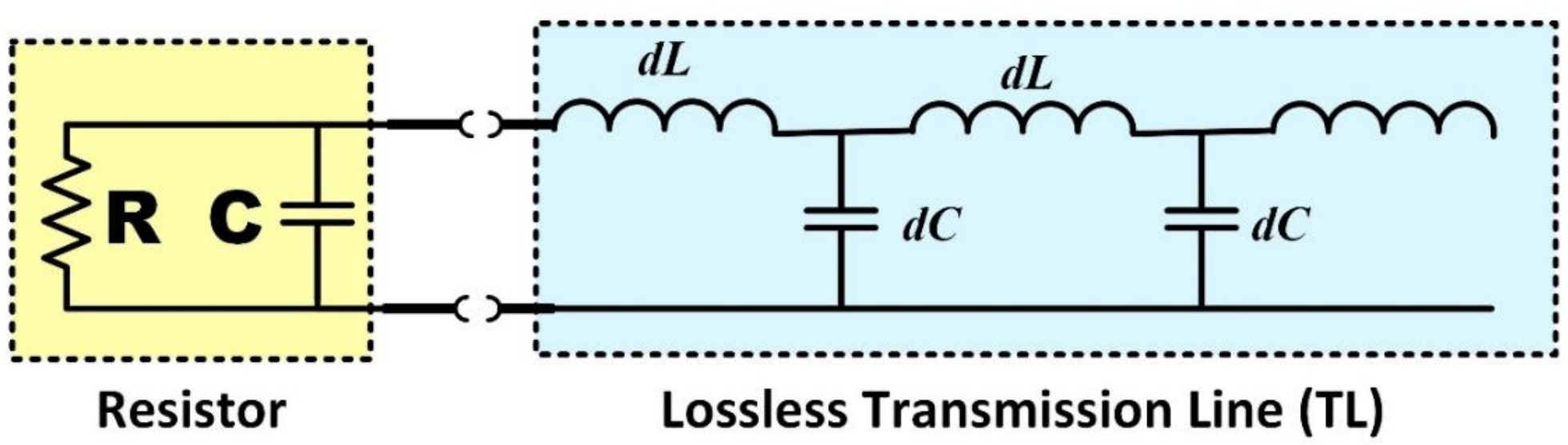

**Fig. 5**. Transmission line replacing the i-v converter of Fig. 4 to explain the meaning of the dielectric relaxation time $\tau_d$ of the material of the resistor (see the text).

To feed our TL with Johnson noise I will design a 50Ω resistor using material of resistivity $\rho=1/\sigma=5\ \Omega\times$cm like silicon doped by $N_d\approx10^{15}$ cm$^{-3}$ donors. A cube of L=1mm side of this material with metal alloyed on two opposed faces acting as terminals, would show $R=\rho\times L/A=50\Omega$. From the dielectric permittivity $\varepsilon>\varepsilon_0$ of Si, $C_{mat}=\varepsilon\times A/L=0{,}1$pF would be its capacitance between these terminals. Since the time constant $R\times C_{mat}=5$ps of this device is: $\tau_d=\varepsilon\times\rho$ (i. e. the Maxwell relaxation time of this material), the voltage step $\Delta V=\pm1.6\ \mu$V caused by each fluctuation would decay with $\tau_d=5$ps. Thus, its Johnson noise would be a Lorentzian density $S_V(f)$ of 4kTR V$^2$/Hz amplitude and $f_c=1/(2\pi\tau_d)\approx32$GHz cutoff frequency. Since $f_c$ and $S_V(f)$ come from applying TEQ to $C_{mat}$, the mean square voltage of its Johnson noise at room T would be: $<(v(t))^2>=kT/C_{mat}=202\ \mu V_{rms}$. This cubic design shows that using "slow" material of high $\tau_d$ in its volume $V_Q$, the capacitance $C_{mat}$ that shunts its R=50Ω will be high.

Working at $f$=32GHz, this 50Ω resistor would not match our TL because its 50Ω resistance shunted by the -j50Ω reactance at $f_c$ of its $C_{mat}$ give: $Z(f_c)=(25\text{-}j25)\ \Omega$. This impedance mismatch between $Z(f_c)$ and $R_0$ gives rise to a situation where 80% of the noise power generated by the resistor at 32 GHz enters the TL and 20% is reflected. Since at $f$=3,2 GHz the reflected power only is 0,25%, I will consider that this resistor matches acceptably our TL up to $f$=3,2 GHz. To get this acceptable matching at 32 GHz we must use faster material like ten times more conductive Si of similar ε to get $\tau_d=\varepsilon\times\rho=0{,}5$ ps. The new $f_c\approx320$ GHz of the Johnson noise of this faster resistor means that its mean square would be √10 times higher: $<(v(t))^2>=640\ \mu V_{rms}$. For Si doped with $N_d\approx10^{16}$ donors/cm$^3$ ($\rho\approx0{,}5\ \Omega\times$cm) a cube of this material with L=1mm only would give R=5Ω between terminals in opposed faces. Keeping A=(1mm×1mm) as the cross section of this "faster" resistor, its length should be L=10mm to give R=50Ω, now shunted by ten times lower $C_{mat}$ between terminals ($C_{mat}=0{,}01$pF, thus giving $\tau_d=RC_{mat}=0{,}5$ps).

Due to its $\tau_d=0.5$ps this faster resistor would match acceptably our TL at $f$=32GHz, where its R=50Ω shunted by –j500Ω due to its $C_{mat}$ at 32 GHz, give $Z=(49{,}5\text{-}j5)\ \Omega$, close enough to $R_0=(50+j0)\ \Omega$ matching perfectly our TL. Effects like skin depth have been left aside to show how the shunting $C_{mat}$ of a resistor decreases as it does the $\tau_d$ of its material, but it never is null. Connecting again our first cubic resistor to our TL, its $C_{mat}=0{,}1$pF would be shunted by $R_t=(R\times R_0)/(R+R_0)=25\Omega$. The Lorentzian noise at the sending end of our TL would be: $4kTR_t=2kTR$ V$^2$/Hz and its $f_c=1/(2\pi R_tC)\approx64$GHz to keep TEQ in its $C_{mat}$. Since the loading of the TL halves the noise voltage $v(t)$ of its $C_{mat}$, its density of voltage noise would be kTR V$^2$/Hz, not V$^2$/Hz as TEQ demands in $C_{mat}$. The lacking 50% is the noise coming from the matched resistor of the SA at the far end. At room T, this 50Ω resistor of the SA would be injecting a density kTR V$^2$/Hz in the TL that would arrive in our cubic, 50Ω resistor. Thus, the 2kTR V$^2$/Hz that keeps TEQ in our resistor is 50% generated in its $C_{mat}$ and 50% in the capacitance C* of the matched resistor of the SA at the far end of the TL.

Similarly, the noise voltage in the matching resistor of the SA would be 50% due to its C* driving its load $R_t=25\Omega$ and the other 50% would be the kTR V$^2$/Hz received from the $C_{mat}$ of our cubic resistor. This sum "in power" keeping TEQ in each resistor is right because the noise collected in our cubic resistor from thermal radiation that pervades its volume $V_Q$ and that collected by C* from a different thermal bath at the far end of the TL are uncorrelated. The 2kTR V$^2$/Hz amplitude found in $C_{mat}$ as $f\rightarrow0$ and its $f_c$=64 GHz give $<v^2(t)>=kT/C_{mat}$ V$^2$, as expected from TEQ. To get this result I have assumed that the matching resistor of the SA was equal to my cubic one, both showing Johnson noise of 2kTR $V^2$/Hz amplitude and $f_c$≈64GHz at the ends of our TL. Now, let me consider the noise we would measure if the 50Ω resistor of the SA was made from Extra-Slow material of $\tau_d$=0.5ms ($10^8$ times 5ps) whose "giant capacitance" $C_{ES}$=0,5ms/50Ω=10μF ($10^8$ times 0,5pF) would shunt its R=50Ω at the input of the SA.

Our TL with its sending end matched by our 50Ω resistor of $\tau_d$=5ps would be seen as a 50Ω load by this extra slow resistor of the SA. This would be so up to $f$=3,2 GHz for the reasons given about its matching to our TL. Thus, our SA would measure the voltage noise of a resistor of $R_t$=25Ω shunted by $C_{ES}$=10μF at its input, whose cutoff frequency is: $f_c=1/(2\pi R_t C_{ES})$=640Hz. Put it bluntly, our SA would show Lorentzian Johnson noise of a 25Ω resistor filtered by a giant $C_{ES}$=10μF. The Johnson noise 2kTR $V^2$/Hz of our Si resistor loaded by our TL only would be flat up to $f_c$=640Hz due to its strong filtering and despising $C_{ES}$ to contend that the Johnson noise of the 25Ω resistor at the input of the SA (i. e. the 2kTR $V^2$/Hz density measured at 6Hz, 60Hz and 200Hz by our SA) remains flat at higher frequencies is wrong. This role of $\tau_d$ by its associated $C_{mat}$ is very important for electronic noise regarding their interaction with its thermal bath [3].

## 4- Discriminating excess noise and shot noise

From the electrical viewpoint, each fluctuation is a Dirac's delta of current whose integration in time by the $C_{mat}$ of a resistor gives its noise voltage *v(t)* shifting discretely by steps of $\pm|q/C_{mat}|$ volts and null risetime each time it occurs. Meanwhile, this voltage *v(t)* decays incessantly with the time constant $\tau=RC_{mat}$ of the natural frequency s=-1/τ of its *i-v conversion gain* $Z_{RC}(j\omega)=1/Y_{RC}(j\omega)$. Thus, the admittance of a resistor shunted by its density $S_{Ish}$=4kT/R $A^2$/Hz of impulsive shot noise, is a proper circuit for its electronic noise. The same holds true for a capacitor of capacitance C (think of its kT/C noise) when it is shunted by an inductor L to form the familiar L-C tank resonator whose losses are considered by a shunting resistance $R_p$ (see below). High losses needing a low $R_p$, entail a high rate λ of translocations of electrons between terminals following Eq. 1. From this series of *h(t)* impulses disturbing the sinusoidal carrier sustained by the electronics in this resonator, we could obtain its random phase modulation [8] (i. e. its phase noise) and our results confirmed perfectly the empirical formula of Leeson [9].

Johnson noise from displacement currents in a resistor, accounting for $P_{TE}$=kT/τ watts that it dissipates in TE, would round-off the *thermal agitation of electric charges* of the titles of [5, 6] by showing the capacitive path that they need to do it. Without its alloyed metal plates, my cubic resistor of previous Section would become a nude cube of Si lacking the path that electrons need to agitate noisily (i. e. a noiseless cube of Si). Thus, let us assign noise to the *device* where it is found and not to its conductive material that can be "none." Think of the voltage noise of two metallic, parallel plates in the vacuum enclosed by a sealed glass bulb, due to electrons agitating in its capacitance shunted by the conductance that thermionic emission from its plates provides at T>0 [10].

Given that the admittance of resistors disguises their discrete shot noise as Johnson noise with Lorentzian spectrum, Nyquist work [6] would be the first report on this shot noise in resistors of density 4kT/R $A^2$/Hz inferred from its effect: the density ATR $V^2$/Hz that Johnson could measure [5]. A *noisy current was inferred* in 1928 from the voltage noise measured as its effect. This inference applies to Hall-effect cells, *i-v* transformers of current clamps, etc. In summary: from voltage measured in a 2TD, we infer the current that has been its cause provided *electric charges do not pile-up between* its terminals.

Since voltmeters give the voltage at each instant of time between two terminals, the lack of such a piling-up grants that the current entering this 2TD and that leaving it are equal. This grants that this voltage comes from a single current whose value can be inferred.

Because in 2004 it was claimed that charges "piling-up" between the terminals of a resistor of high $\tau_d$ produced shot noise from its bias currents [4], let me analyze this idea after saying that inspired by [6], Callen and Welton had published in 1951 a work [7] that I met in 2008 from my interaction with authors of [4], who had measured noise in resistors of CdTe of $\tau_d \approx 0.3$ ms, where $\tau_d$ was much higher than their transit time for electrons $\tau_t$. Under this $\tau_d >> \tau_t$ condition, they expected to find that electrons in transit of a current $I_C$ biasing this resistor, would "pile-up" somehow between its terminals. In their own words: "*charge neutrality of the device can be violated and shot noise can be observed*" [4]. Since shot noise coming from a bias current never has been found, this conjecture was worth studying after discarding $\Delta R$, the undeniable resistance noise that their noise data, evolving as excess noise $\Delta R \times (I_C)^2$, were showing unambiguously.

Excess noise in a resistor biased by a current $I_C$ is noise voltage emerging over its Johnson noise (floor), hence its name. Its spectral density proportional to $(I_C)^2$ means that it comes from resistance noise of density $\Delta R$ ($\Omega^2$/Hz) converted by $I_C$ into noise voltage of density $\Delta R \times (I_C)^2$ ($V^2$/Hz). A paradigmatic example of this conversion is Fig. 2 of [4] that their authors present as a spectral density of fluctuations of current in their device. For $I_C=0$ it should be its Nyquist noise $S_I(f)=4kT/R$ $A^2$/Hz, as if it was converted into voltage out of its $C_{mat}$, by OAs working as described by Fig. 4. Reading CdTe of $\tau_d \approx 0.3$ ms, I asked these authors for their reasons to despise $C_{mat}=\tau_d/R=1{,}3$pF (and $C_{stray} \approx 0{,}5$ pF that they gave in ref. [16] of [4] for their setup). Since I had explained flicker noise in vacuum devices [10] and 1/*f* excess noise in solid-state ones [11] from noise generated in capacitors, I wanted to know their reasons to despise them.

Readers used to electronic feedback will know that the red capacitance of Fig. 4 weakens the negative feedback of its OA as *f* increases. Due to this weakening, a noise of the OA (its density $e_n$ referred at its input) is amplified more as *f* increases. This adds to its $V_0$ a spurious noise term $S_{OA}(f)$ that increases with *f*. People not having dealt with shunting capacitance could read ref. [16] of [4], to find why the capacitance shunting the resistor bends upwards the otherwise flat density considered $S_I(f)=4kT/R$ $A^2$/Hz at the output of the OA for $I_C=0$. Anyhow, $C_{mat}$ was despised for these devices of R=233MΩ whose $\tau_d=0.3$ ms has inspired $\tau_d \approx 0{,}5$ ms linked with $C_{ES}=10\mu F$ in Section 3. For $f \gtrsim 10$kHz, the term $S_{OA}(f)$ disturbs so much the otherwise flat noise voltage at the output of the OA that no measurements over 10 kHz were shown in [4], where it is written:

*"Current noise experiments have been performed on the CdTe resistor by means of the correlation technique implemented on a state of the art noise spectrum analyzer able to probe noise levels as low as $10^{-30}$ $A^2$/Hz and to reach frequencies up to $10^5$ Hz at room temperature [16]. The characteristics of the measurement setup allow reaching the extremely low current noise levels present in semi-insulating materials and to cover a wide enough range of frequencies to get rid of the 1/f contribution that may hide the presence of the shot noise plateau."*

To get rid of the 1/*f* contribution, noise data from $10^4$ to $10^5$ Hz, much less affected by such contribution than those ending at $10^4$ Hz, should be in Fig. 2 of [4]. But even this detail fades away given the big drawback regarding the 1/*f* contribution of this figure. As it is well known, $1/f$ excess noise comes from resistance noise converted into noise voltage by $I_C$ biasing the resistor [12, 11]. Thus, the density (in $V^2$/Hz) of this converted voltage increases as $(I_C)^2$ because each fluctuation of resistance ($\Delta R$ in $\Omega$) is converted into $\Delta R \times I_C$ volts. In this way, a spectral density $S_R(f)$ $\Omega^2$/Hz gives rise to $S_{VR}(f)=(I_C)^2 \times S_R(f)$ $V^2$/Hz, a proportional density of excess noise that emerges from (or surpasses neatly) the 4kTR $V^2$/Hz floor that the Johnson noise of the resistor is for excess noise, see our Fig. 2. The drawback of [4] is that all the noise of its Fig. 2, being proportional to $(I_C)^2$, reveals an overwhelming resistance noise $\Delta R$ that should have been considered.

Taking the first curve of Fig. 2 in [4] for $I_C$=7.9nA, the second one for $I_C$=23.7nA runs parallel at ≈9.5dB over the former for frequencies between 10Hz-5kHz approaching three decades. Since $(23.7\text{nA}/7.9\text{nA})^2$=3 and 9,5dB means nine times more noise power at each $f$, a resistance noise $\Delta R$ seems to generate all the noise data shown here by all its curves being proportional to $(I_C)^2$. The third one for $I_C$=67nA running at ≈9dB over that of $I_C$=23.7nA, gives: 67/23.7=2,83. Thus, we find again $2{,}83^2$=7,99 being equal to the power ratio 9dB=7,94. This evolution with $(I_C)^2$ that holds true for all the curves up to $I_C$=194nA justifies my feeling on an overwhelming resistance noise $S_R(f)$ $\Omega^2$/Hz that each $I_C$ would reveal as excess noise $(I_C)^2 \times S_R(f)$ in this device. Although the unveiling of $S_R(f)$ by $I_C$ [12] should be well-known in 2008, my comment on [4] was taken with reluctance to question in 2008 this work published in 2004.

The answer of their authors using "the Fluctuation-Dissipation Theorem" led me to [7], where I could see that the admittance $Y_{RC}(j\omega)$ that I was using for electronic noise was founded. To store the fluctuating energy observed as fluctuating voltage, one needs capacitance to have electronic noise accordingly to [7]. Thus, the capacitance $C_{mat}$ of resistors was right as well as those of the floating gates of [10, 11]. Handling fluctuations and dissipation of energy in noisy devices demands a complex immittance. Let me say, however, that the generic impedance of [7] (a series notion) is not the clever choice for Johnson noise (a voltage). Electromagnetic energy interacting with matter by dissipative (i. e. conduction) and non-dissipative (i. e. displacement) densities of current, suggests that the parallel notion Admittance is the clever choice. Fig. 1 of [3], devoid of parasitic inductances that could hide this meaning, illustrates this choice.

Studying the noise voltage of a resistor in Fourier domain, dissipation of electrical energy demands resistance $R(\omega)$ or conductance $G(\omega)$ in its noise circuit to have sinusoids of voltage $v(t)$ and current $i(t)$ in phase whose product at each instant of time $p(t)=v(t)\times i(t)$ is the non-negative function of Fig. 6 called *active power*. Each *i-v* pair with this phase condition means electrical energy entering this 2TD to become heat, thus *dissipation*. By contrast, electrical energy that enter or exits the resistor while keeping its form (i. e. that *fluctuates* in it) requires sinusoids $v(t)$ and $i(t)$ in quadrature (±90º-off whose product at each instant of time $p(t)=v(t)\times i(t)$ is the zero-mean oscillation of Fig. 7 called *reactive power*. Since fluctuations entail reactive power, reactance $X(\omega)$ is found in a series model of the resistor by its Impedance $Z(j\omega)=R(\omega)+jX(\omega)$ where $\omega=2\pi f$ or susceptance $B(\omega)$ is needed for its parallel modelling by its admittance $Y_{RC}(j\omega)=G(\omega)+jB(\omega)$ [2, 3].

$Z(j\omega)$ appearing in [7] shows the need for a complex immitance to represent a noisy device, where reactance handles fluctuations of energy in time that resistance is unable to handle. By contrast, resistance handles dissipation that reactance cannot handle and when these two magnitudes are properly combined, they give results with physical sense like the impulse responses $h(t)$ (device reactions). Each $h(t)$ shows the one-way character of the noisy process (i. e. its cause-effect dynamic) clearly illustrated by those charge relaxations in resistors [3] agreeing with the *Irreversibility and generalized noise* entitling [7]. Dissipation cannot build in time an instantaneous unbalance of energy like a fluctuation nor the effect of a sudden fluctuation can influence its cause, already gone. In our model, fluctuations involve a single electron and are instantaneous, but dissipating energy of $C_{mat}$ by currents driven by its own noise voltage *v(t)*, takes time to occur and involves all the carriers of the resistor by the field *v(t)/d* that they perceive.

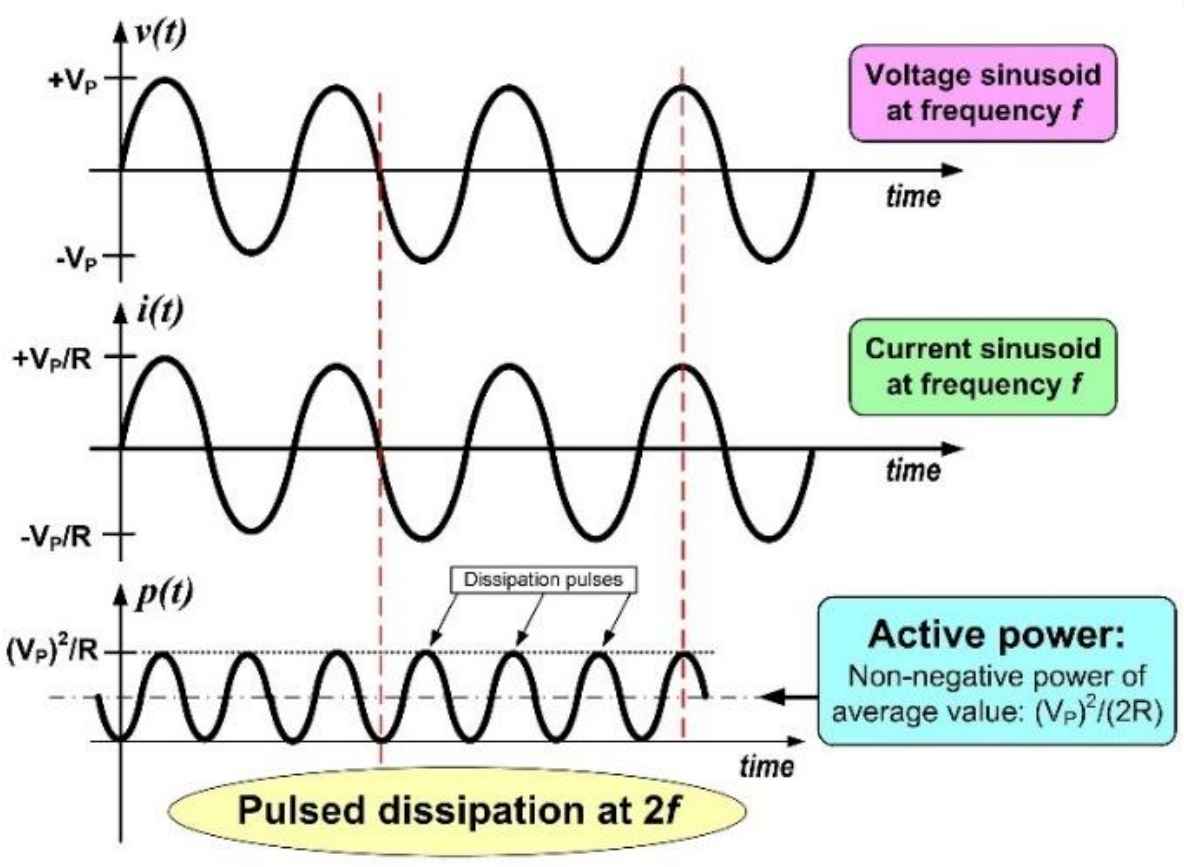

**Fig. 6**. Active power coming from the product at each instant of current and voltage sinusoids of the same frequency found in phase in a 2TD, thus being a dissipative device. (see the text).

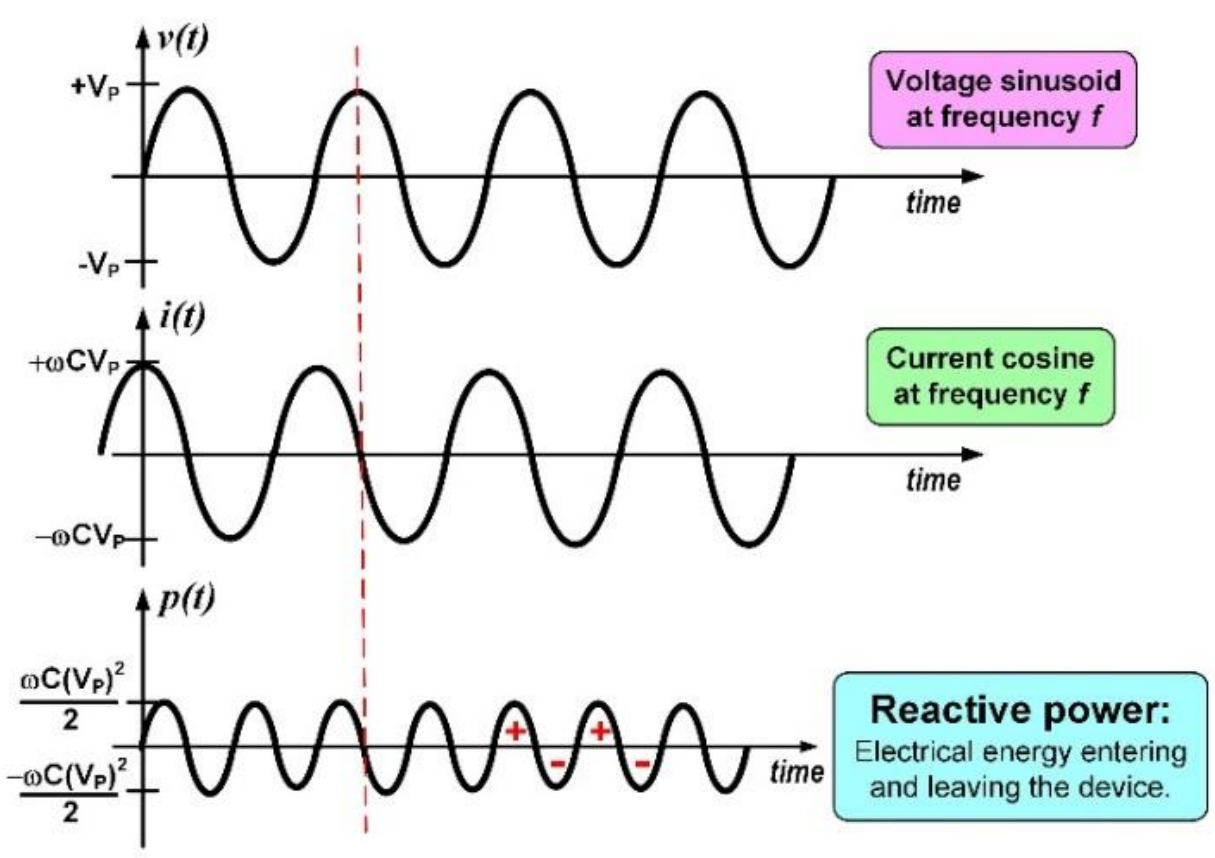

**Fig. 7**. Reactive power coming from the product at each instant of current and voltage sinusoids of the same frequency found in quadrature in a 2TD undergoing fluctuations of its electrical energy.

Given the attention paid in [4] to its material and the scarce interest on the device that generates noise, I will say that studying noise requires a source of it like a 2TD, not a piece of material. For a resistor having metallic gaskets at the ends of a rod of n-type silicon, these terminals allow applying electric fields to its rod to have a *load* dissipating energy and they also allow collecting its voltage and currents to have an electrical *source* of energy. Fluctuations of electric field between its terminals and charge displacements between them are linked by their capacitance. We will find noise in this Si rod if we clad it by two gaskets to form a leaky capacitor whose capacitance allows states of energy whose thermal occupation give rise to Johnson noise. Capacitance allows this resistor to interact with its thermal bath [3] and gives shelter to the fluctuating energy that gives rise to the varying voltage that drives incessantly this dissipation.

As soon as two metallic objects are at distance $d$ in space, their capacitance C is there allowing electrons to pass between them as displacement currents while thermionic emission shunts their C by a conductance G [11] to form the complex admittance $Y_{RC}(j\omega)$ of a noisy 2TD. Next Section considers states of energy for electrons in a 1MΩ resistor due to its $C_{mat}=\tau_d/R$. For n-type Si as conductive material, I will consider the carriers that allow setting electrical current in an n-type Si bulk region between two ohmic contacts (terminals). I mean electrons with energy to be in the conduction band of this material, thus "*conductive electrons*." The current $i(t)=v(t)/R$ that they make possible in our device is dissipative current that obeys Ohm's Law in dc, thus "conduction" current.

## 5- Electronic noise allows assigning shot noise

Considering my 1MΩ resistor with its two terminals (ohmic contacts) plenty of conductive electrons, let me begin with the instant translocation of one of them from terminal A to terminal B when its $C_{mat}$ was discharged. To appear translocated in terminal B, the electron of terminal A would absorb $E_1=q^2/(2C_{mat})$ joules. This fluctuation leaving charges $+q$ and $-q$ in its terminals, would set instantly a voltage $\Delta V=V_A-V_B=q/C_{mat}$ volts between them. From the energy set in $C_{mat}$ by this fluctuation, two displaced electrons in $C_{mat}$ would set $E_2=2^2\times E_1$ and so on. However, these states of energy $E_n>E_1$, with several electrons displaced in $C_{mat}$ (useful to study resistors made from materials of high $\tau_d$) fall out of our scope seeking to grant the $\lambda=3{,}21\times10^{11}$ fluctuations per second of Eq. (1) for our 1MΩ resistor at room T. For this to be so, its inner material must have $\tau_d$ below $T_{avg}$ ($\tau_d\leq T_{avg}=3{,}1$ ps) taking $T_{avg}=1/\lambda$ as an average time between its random fluctuations.

For $\tau_d=T_{avg}$, the cutoff frequency of its Johnson noise would be $f_c=51$ GHz and more than 86% of $E_1$ would be removed from its $C_{mat}=\tau_d/1M\Omega=3{,}1\times10^{-18}$ F during $T_{avg}$ by its R=1MΩ. From the $E_1$ of each $h(t)$ decaying down to $0{,}14E_1$ ($1/e^2=0{,}14$) and given that fluctuations with opposed sign to the precedent one should appear very often, a good removal of $E_1$ seems granted. Thus, I will consider material of $\tau_d\approx3{,}1$ ps as fast enough to get rid of the energy $E_1$ during $T_{avg}$. This $\tau_d$ is close to $\tau_d\approx5$ ps of a resistor made from lightly doped Si ($N_d=10^{15}$ cm$^{-3}$) in Section 3. Watching the shape of our new 1MΩ resistor and its low $C_{mat}$ (considering valid the 1-D model of $C_{mat}$ and R for its Si rod) one finds that its rod should be a *wire* of L=1cm and 19 microns diameter. This filamentous form strongly suggests that the capacitance that allows the discrete dissipation of Eq. (1) is a *microscopic* one. Before studying this possibility let me complete my reasoning about my 1MΩ resistor made from n-Si of $\tau_d=T_{avg}$. From Eq. (1) and using $\tau_d=1/\lambda$ we obtain:

$$\tau_d = RC_{mat} = \frac{1}{\lambda} = \frac{q^2R}{2kT} \Rightarrow C_{mat} = \frac{q^2}{2kT} = \frac{1}{2}\frac{q}{V_T} = \frac{1}{2}C_f \qquad (2)$$

The striking change of $C_{mat}=\tau_d/R$ (the macroscopic capacitance coming from the dielectric nature of the material between terminals) by a "thermal capacitance" given by the ratio between $q$ and the thermal voltage $V_T$, reinforces the above suggestion. I mean that the fast removal of capacitive energy between terminals requires a low capacitance, close to the tiny one $C_f=q/V_T$ that I met years ago while studying donors in AlGaAs. Since this leads to consider how energy stored capacitively becomes heat released to the atom lattice that is the subject of the Annex, I will end this part giving two relevant results of our model that do not need the Annex yet. Going back to my cubic resistor of Section 3, let me consider the power $P_Q$ of *charge noise* (in C$^2$/s) that exists in its $C_{mat}=0{,}1$ pF. Half this power will be its λ fluctuations per second ($P_{Qf}$) and the other 50% ($P_{Qs}$) will be its λ slower recoils $i_d(t)$ per second, decaying with time constant $\tau_d$ as they displace-back each electron that was displaced as a fluctuation. Due to their cause-effect connection, $P_{Qf}$ and $P_{Qs}$ do not overlap in time, thus adding in power as they are in Eq. (8) of [3], which is:

$$P_Q = P_{Qf} + P_{Qs} = \lambda q^2 + \lambda q^2 = 2\lambda q^2 = 2\frac{2kT}{q^2 R} q^2 = \frac{4kT}{R}\ {C^2}/{s} \qquad (3)$$

This charge noise power in a resistor at temperature T is a number in $C^2/s$ that equals its flat Nyquist density $S_I$ in $A^2/Hz$. Since $P_{Qf}$ is a series of Dirac's deltas from the electrical viewpoint, it is an impulsive current that we must infer from its effects. Due to $Y_{RC}(j\omega)$ the effects of these deltas appearing randomly and with random sign, is a series of voltage shifts of $\Delta V=\pm q/C_{mat}$ volts between terminals that, decaying with time constant $\tau=RC_{mat}$, form its Johnson noise $v(t)$. Hence, differentiating the $v(t)$ of a resistor we could know the impulsive current corresponding to the fast term $P_{Qf}$. At room T, however, this method would require "ultrafast electronics" able to count positive and negative pulses at the huge rate $\lambda$ set by Eq. (1) for $R=50\Omega$ of this cubic resistor.

However, low-speed electronics measuring its Johnson noise $v(t)$ at low $f$ can give us the rate $\lambda$ because we know that each charge displaced-back between terminals during each $h(t)$ decay of voltage is $q$. Discrete charges of $q$ coulombs each, passing with mean rate $\lambda$ between terminals, entail a shot noise density $S_{ish}=2q^2\lambda$ $A^2/Hz$ as $f\rightarrow 0$ that $Y_{RC}(j\omega)$ will convert into a density of noise voltage $S_V=S_{ish}\times R^2$ $V^2/Hz$, which is easy to measure. Since each recoil $i_d(t)=C\times(dv(t)/dt)$ between terminals is the "slow" return of the charge $-q$ displaced instantly in opposed sense by each fluctuation, the assignment of shot noise to these recoils of $P_{Qs}$ is well founded, and their shot noise converted into a noise voltage density ($V^2/Hz$) by $Y_{RC}(j\omega)$ will appear in our experiments. Since this random series of recoils entails the passage of $\lambda$ electrons per second independently one of each other, their shot noise at low $f$ will be: $S_{Ish}=2q^2\lambda$ $A^2/Hz$. From Eq. (1) we have:

$$S_{Ish} = 2q^2\lambda = 2\frac{2kT}{q^2 R} q^2 = \frac{4kT}{R}\ {A^2}/{Hz} \qquad (4)$$

Taking $R^2$ times Eq. (4), the voltage density measured at low $f$ will be:

$$S_V = \frac{4kT}{R} \times R^2 = 4kTR\ {V^2}/{Hz} \qquad (5)$$

Eq. (4) shows the impulsive origin of the Nyquist noise [2, 3] and Eq. (5) gives the amplitude of the Lorentzian spectrum of Johnson noise that it should give. For the cubic resistor of Section 3, its $\tau_d$=5ps would be the time constant of its recoils $i_d(t)$ that would form its $P_{Qs}$, see Fig. 4 of [3]. Eqs. (4) and (5) showing that the Johnson noise of the resistor is accounted for by those recoils $i_d(t)$, give two relevant results. The first one is the impulsive, random current whose spectral density is the Nyquist noise inferred from the Johnson noise we can measure. The second result is that whereas each displacement current $i_d(t)$ generates noise the simultaneous dissipative current $v(t)/R=-i_d(t)$ that occurs as charge relaxes in the resistor [3], *does not add noise voltage at all.*

If the Johnson noise of a resistor comes from the shot noise of its recoils of charge $i_d(t)=C\times(dv(t)/dt)$ caused by its fluctuations, assigning shot noise to its simultaneous dissipative currents $v(t)/R=-i_d(t)$ is unfounded from our model for electronic noise. And if the inability of currents driven by *v(t)* to generate noise holds for a cubic resistor of 1mm side and R=50Ω, it will hold for any resistor made from two cubic resistors put in series (R=100Ω) or in parallel (R=25Ω). Using this idea we can show that the inability of currents driven by the voltage *v(t)* to give noise, holds for any resistance value. Thinking of resistors of very high R, their Johnson noise will reach hundredths of millivolts. If the inability of dissipative currents to generate noise in a resistor holds for hundredths of mV, bias currents $I_C=V_{DC}/R$ entailing similar $V_{DC}$ voltages should be noiseless too.

From the linear ratio $R=V_{DC}/I_C$ of resistors up to $V_{DC}$ values well over those of their Johnson noise (provided heating effects are negligible), or from the reasoning of previous paragraph combining noiseless resistors in series and in parallel to reach any voltage or current, the inability of conduction currents to generate noise should be true for any $V_{DC}$ value. Hence, shot noise from bias currents as claimed in [4] is hard to find following our model, and the lack of shot noise for $I_C \approx 0{,}7\mu A$ in our resistor of Fig. 1 no longer is the puzzling result that it was up to now, but the expected proof of its null noise predicted from our model. Concerning the power dissipated in this resistor by its $I_C$, it is: $P_{Dis}=1M\Omega \times (0{,}7\mu A)^2 \approx 0{,}5\mu W$, whose heating effects seem negligible for these devices of macroscopic size (see Fig. 8). This is why for $I_C=0{,}7\mu A$ and for $I_C=0$ in Fig. 2, its noise voltages look equal within experimental error down to $f$=100Hz, where $I_C=0{,}7\mu A$ starts to reveal the excess noise of our resistors as $f \to 0$ [11, 12].

It is worth noting that our 1MΩ resistor (DUT) in Fig. 1 is in TE at room T and it is shunted by the $C_{in}/2$=7,5 pF of the balanced input of the EGG&PAR 113 amplifier in *differential mode*. This capacitance comes from the two unbalanced inputs of this LNA ($C_{in}$=15 pF each) working in series to shunt the DUT. This stray capacitance must be increased by the wiring capacitance of the two coaxial cables connecting each terminal of our DUT to each input of our LNA. Since the length of each RG-58 coaxial cable is ≈14cm, its capacitance will be close to $C_{coax}$=16pF, thus adding 8pF to $C_{in}/2$. Therefore, our DUT is a resistor of $R_t$=1MΩ shunted by $C_t \approx$16pF that form a first order, low-pass filter of cutoff frequency $f_{C3}=1/(2\pi R_t C_t)$. This is the third antialiasing filter mentioned in the footnote of Fig. 1 whose cutoff frequency $f_{C3}$=9,9 kHz is very interesting because it sets the power dissipated by our DUT in TE to $P_{TE}=kT/(R_t C_t)$.

This small power $P_{TE}=2{,}6\times 10^{-16}$ W makes Fig. 2 "*spectacular*" because it shows that increasing the dissipation of my resistor up to $P_{Dis} \approx 0{,}5\mu W$ (thus by $2\times 10^9$ times, 93dB) has been done by its $I_C \approx 0{,}7\mu A$ *without varying its Johnson noise*. For this noise coming from a fluctuation-dissipation dynamic [2, 3, 7], this fact shown by Fig. 2 strongly suggests that conduction current must be totally different from charges passing between terminals like corpuscle-like electrons drifting between them. Guided by this conviction I proposed the model of the Annex for noiseless currents that obey Ohm's Law in dc.

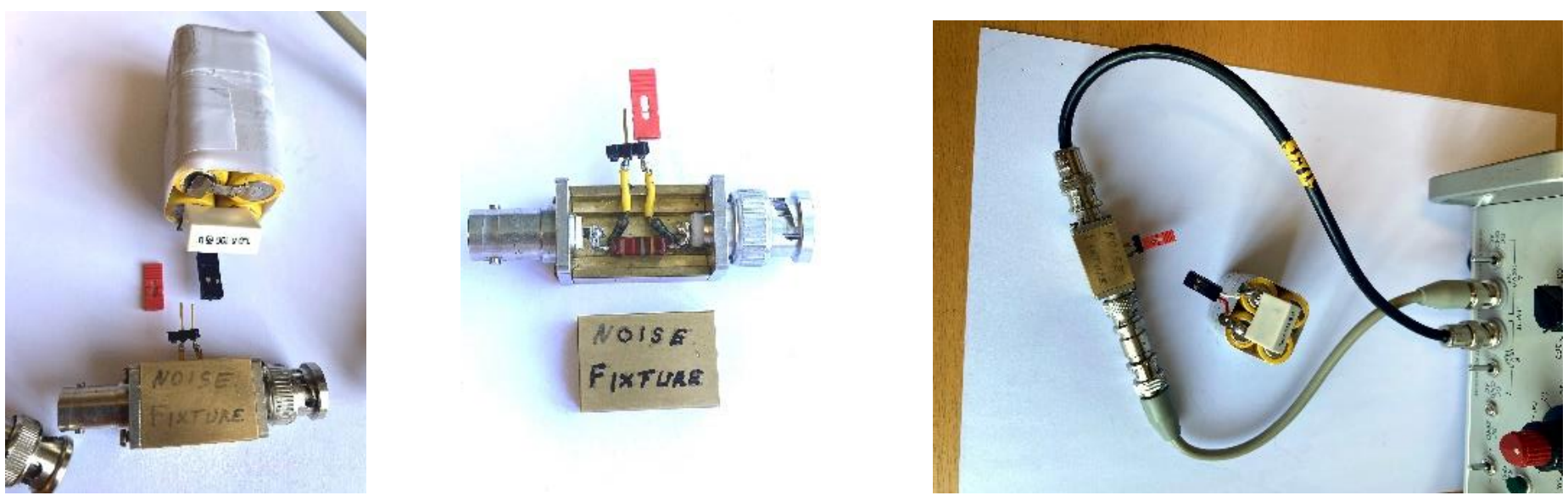

**Fig. 8**. Photographs of the resistors, jumper, battery and setup represented by the circuit of Fig. 1.

## Conclusions

From a model within the quantum framework of Callen and Welton, Johnson noise would come from thermal energy that resistors collect and convert into voltage by their capacitance between terminals. Thus, we must use their complex admittance giving room for their fluctuations of electrical energy and its incessant dissipation, both required to generate this noise accordingly to quantum physics.

Because impulsive currents due to displacements of single electrons between the terminals of resistors (i. e. fluctuations) accounts for their Johnson noise, the dissipative currents they have both in thermal equilibrium or externally biased, should be noiseless. This suggests that assigning shot noise to their bias currents is unfounded.

Because current obeying Ohm's Law in dc would be dissipative current devoid of passages of charge between terminals of resistors other than those fluctuations that give rise to their Johnson noise, a proper model for conduction current is needed. That of the Annex could is given as a starting point.

## Annex: Dissipative current by carriers of energy.

In this work we have shown that assigning shot noise to displacement currents but not to conduction ones, accounts for the Johnson noise of resistors. This motif to lose hope of finding noise from a bias current can be accepted better if one sees how dissipative currents could exist in resistors without disturbing their displacement currents causing Johnson noise. Electrical current departing from charged monopoles in transit between two terminals would reinforce the distinct roles played by the two orthogonal currents that Y(*jω*) demands in a resistor to generate its Johnson noise *v(t)*. Since charges passing between its terminals entail shot noise that would affect *v(t)*, dissipation should occur without displacing charges in $V_Q$ other than its thermal fluctuations. Thus, I will propose a dissipative process preserving its fluctuations in TE, guided by the capacitance $C_f=q/V_T$ of Eq. (2) and this phrase about Eq. (1) suggesting that "*...its conductance G has to do with its ability to dissipate energy λ times per second too.*"

The mean power $P_{DC}$ dissipated in a resistor of R ohms biased by a current $I_C$ is:

$$P_{DC} = R \times I_C^2 = \frac{(R \times I_C)^2}{R} = G \times V^2 \qquad (6)$$

If the tiny (likely microscopic) capacitance $C_f$ of Eq. (2) was able to sense this voltage V between terminals of the resistor, the energy $U_{joul}$ that it would be storing is:

$$U_{joul} = \frac{1}{2} C_f \times V^2 = \frac{1}{2} \frac{q}{V_T} \times V^2 = \frac{q^2}{2kT} \times V^2 \qquad (7)$$

Now, let us think of switched capacitor circuits dissipating energy without needing resistors with sinusoidal current and voltage in-phase. Loading $U_{joul}$ in $C_f$ and shorting next its terminals briefly, $U_{joul}$ would lose its electrical form to become electromagnetic energy radiated to its environment and some heat to obey Thermodynamics. Repeating m times per second this process, we would dissipate $P_{avg}=m \times U_{joul}$ watts by this pulsed form that, if m was high enough, could emulate a continuous dissipation of mean value $P_{DC}$. Dividing $P_{DC}$ of Eq. (6) by $U_{joul}$ of Eq. (7), the average rate of packets $U_{joul}$ that our resistor would be dissipating by its conductance G=1/R would be:

$$\frac{P_{DC}}{U_{joul}} = \frac{2kT}{Rq^2} = \frac{2kT}{q^2} G = \lambda \qquad (8)$$

Eq. (8) shows that taking the conductance G=1/R of a resistor as "*its ability to dissipate λ packets per second of capacitive energy*" paves the way for the dissipative current that I am looking for. I mean current increasing the dissipation of a resistor by orders of magnitude (e. g. 93dB) without affecting its Johnson noise, but releasing to the lattice as heat the energy $U_{joul}$ each time a fluctuation occurs. About $C_f$, I met it in the eighties while studying donor atoms giving conductive electrons in AlGaAs (DX centers). Being used to Si atoms in n-AlGaAs let me think of a Si atom in AlGaAs as the spherical dipole that its $Si^+$ cation of +*q* C form with the cloud of charge -*q* C of its outer electron. This dipole exists when this electronic cloud is trapped as a sphere of charge -*q* C around the $Si^+$ cation in its center, whose mean radius $d_0$ would give "a size" for this donor trap.

Bound in this way to its $Si^+$ cation, this outer electron does not contribute to the G=1/R of a macroscopic AlGaAs resistor. For usual voltages V between terminals at distance $d \gg d_0$, the fraction of V affecting this dipole ($\approx V \times d_0/d$) would not break it. However, this dipole of charges $+q$ and $-q$ at distance $d_0$ would sense and react to electric fields of its thermal bath, exchanging energy that it can store by increasing its mean distance that would be $d_0+d(T)$ at each T. Since its two "plates" have constant charges $+q$ and $-q$, their $d_0+d(T)$ would increase with T. By contrast, their capacitance C(T) would decrease as T increases, thus increasing the mean thermal energy $U_{sph}(T)=q^2/(2C(T))$ stored in this dipole as T increases. This $U_{sph}$ and the mean volume of the outer electron enclosing its $Si^+$ cation would keep rising with T up to a point where the volume of this cloud of negative charge becomes, suddenly, the whole volume $V_Q$ of the device.

When this occurs, the donor atom is ionized and its outer electron no longer is trapped around its $Si^+$ cation. It is evenly distributed within $V_Q$, with Bloch functions in its wavefunction for the periodic medium where it is: the atomic lattice that fills $V_Q$. Now, this electronic cloud of density $-q/V_Q$ C/cm$^3$ is trapped in the whole volume of the resistor between its two terminals, where it has access to both contacts to *sense all the voltage* V between them. In this way it could react to electric fields set in $V_Q$ both by thermal activity ($E(t)=v(t)/d$) or externally ($E=V/d$) by a bias current $I_C$ producing a dc voltage $V=R \times I_C$ between terminals. To measure noise voltages like excess noise and its floor (the Johnson noise of the resistor) both in $V^2$/Hz, you inject a current $I_C$ to allow the appearance of varying voltage to be measured (i. e. you do not apply a constant voltage V).

Resuming my reasoning about the cloud of density $-q/V_Q$ C/cm$^3$ in $V_Q$, ready to react to an electric field $E=V/d$ in this volume, let me guess its possible reaction to this stimulus. To do it, I need first a reason for its uniform distribution in $V_Q$. For a movable cloud of negative charge, a positive charge $+q$ distributed uniformly in $V_Q$ is a good one. This is what I will call hereafter the *positive counterpart* (PC) of a conductive electron in the conduction band (CB) that I used to overlook to think of this "carrier in the CB" as a corpuscular charge of $-q$ coulombs. Hereafter, I will consider that a "carrier" in n-type semiconductors would entail an extended dipole of charges $-q$ and $+q$ distributed in $V_Q$, screening one to each other as best as possible. This differs from a corpuscle-like electron able to travel $d$ in a transit time $\tau_t$ under the action of $E=V/d$. Regarding degrees of freedom for the electronic cloud, it was trapped in a small volume around a $Si^+$ cation at low T. Gaining energy it passes to exist in $V_Q$, a bigger trap based on the force between its charge $-q$ C and a charge $+q$ C that must exist in $V_Q$ from charge neutrality.

I mean the PC of the movable cloud of charge of each carrier, very different from a $Si^+$ cation highly located somewhere in $V_Q$. To force this cloud to be evenly distributed in $V_Q$, the PC that I can imagine should be positive charge distributed in the lattice with density $+q/V_Q$. This could keep the cloud delocalized with a density $\rho_{cl}=-q/V_Q$ C/cm$^3$ everywhere. The reason why this uniform PC of $\rho^+=+q/V_Q$ C/cm$^3$ exists in $V_Q$, or why it is sensed uniformly by the negative cloud of a carrier falls out of the scope of this paper. The negative cloud could "feel in this way" all the $Si^+$ cations that it could use to form neutral Si atoms. The point is that this PC is required to keep local neutrality of charge at the atomic level and a global one in $V_Q$. I will assume too that the electronic cloud of a conductive electron and its PC do not collapse together for those reasons that avoid the collapse of this cloud with its cation when it is the outer electron of a Si atom.

All in all, the idea of an electron in the CB as a carrier of charge -*q* drifting under E=V/*d*, is replaced by this one of an extended dipole of charge in $V_Q$ able to react to E=V/*d*. Readers believing in monopoles of charge could consider that a point-charge always has its PC around by its radial electric field going to (or coming from) infinite. For a point-like electron, this field emulates its PC of charge +*q* distributed on the inner surface of a surrounding sphere. Unable to reach it in an infinite universe, I can think of electrical monopoles as extended spherical dipoles. When the available "universe" is the finite volume $V_Q$ of a resistor where this electron must be, its uniform PC is a simplified way to avoid calculations considering the actual shape of $V_Q$, possible charge (trapped and non-trapped) at its surfaces, etc.

From the above notions let me consider a conductive electron as a cloud of density $\rho^-=-q/V_Q$ C/cm$^3$ in $V_Q$ that screens as best as possible its fixed PC of $\rho^+=q/V_Q$ C/cm$^3$. Because this PC should be a fine-grain density at the atomic level, each atom of the lattice should exhibit a small density of positive charge, binding to it the movable cloud. Thermal activity, that once separated this cloud of charge -*q* C from the $Si^+$ cation where it comes from, would try to separate it from its PC. This suggests how this dipole could store thermal energy: by increasing the mean separation S(T) of its charges as T increases. A carrier in the bottom of the CB would have the lowest S(T) value. The mean S(T) of a carrier at a given T would come from applying thermal equipartition (TEQ) to obtain the average capacitance $\langle C_f \rangle$ between these charges +*q* and –*q* storing kT/2 joules on average for this degree of freedom allowed to electrons. The $\langle C_f \rangle$ thus obtained is:

$$U_f = \frac{q^2}{2\langle C_f\rangle} = \frac{kT}{2} \Rightarrow \langle C_f\rangle = C_f = \frac{q^2}{kT} = \frac{q}{V_T} \qquad (9)$$

Eq. (6) shows how $C_f$ was found studying DX centers in the eighties. In summary: each conductive electron of a resistor would react as a small capacitor of $C_f=q/V_T$ Farads regarding its voltage *v(t)* between terminals. In this way, rapidly fluctuating energy (an "ac term" of thermal origin), would "build" this thermal capacitor for each carrier, thus preparing it to sense other voltages like a dc voltage V superimposed to its Johnson noise. Let me recall that to find Eq. (1) and $C_f$ in Eq. (2), all the capacitive energy set between terminals by each fluctuation was dissipated before the arrival of the next one. Viewing how λ becomes independent of $C_{mat}$ in Eq. (1), I thought that $C_{mat}$ allowing fluctuations to enter the resistor and holding them as energy between terminals makes sense. Now, the point is that Eq. (1) also gives λ as the rate of chances to grant the dissipation of each fluctuation by a huge number of small packets of energy, one for each fluctuation.

Considering that the electrical energy of each fluctuation is removed by the incessant dissipation involving all the carriers of the device, all the conductive electrons between terminals in our resistor should help with the task of this removal. A fast way to do it would be to load energy by their $C_f$ sensing the voltage *v(t)* between terminals at each instant of time to store it in the atom lattice as elastic energy. Note that this is a first step towards electrical energy changing its form (dissipation) which would begin by loading a small energy $U_{Joul}$ in the $C_f$ of each carrier by displacing its cloud of charge from its mean position around each atom of the lattice, that is: from its best screening of its PC at mean distance S(T). This is a polarization of each carrier by V=R×$I_C$ that Figs. 9-a and 9-b try to sketch.

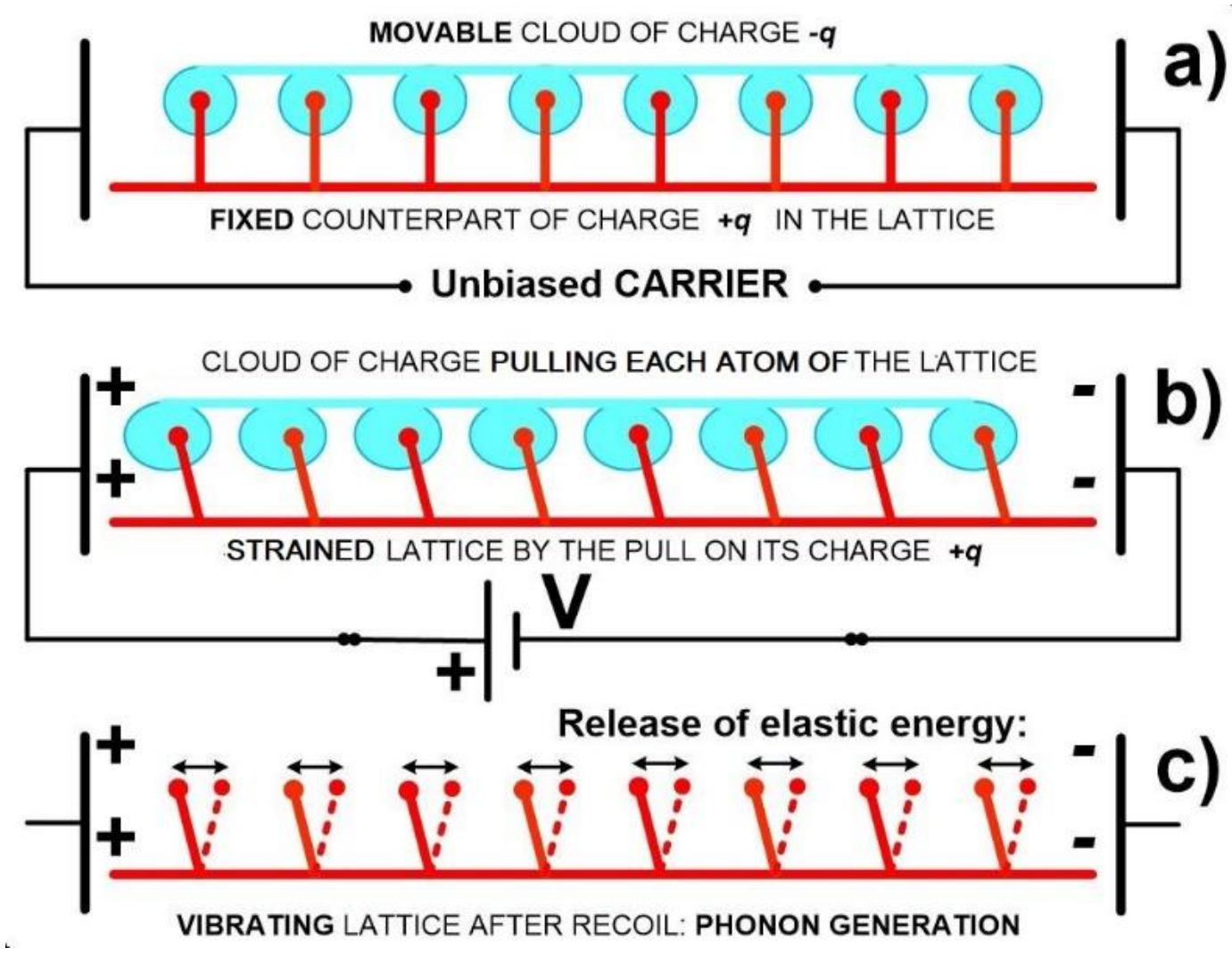

**Figure 9**. Dipolar structure for an electron in the conduction band (carrier): **a)** unbiased, **b)** under bias and **c)** vibrating lattice once released by a biased carrier that disappears.

In the resistor thus biased, the field E=V/*d* would pull towards the anode the cloud of charge of each carrier while its PC in the lattice would be pushed with equal force in opposed sense. This would give a null net force on the resistor containing this carrier. However, these two forces stretching its dipole would produce a small strain of the atom lattice that suggests how dissipation would end. As Fig. 9-b shows, the cloud attracted by the anode would pull each atom of the lattice towards the anode, thus displacing them slightly from their mean positions in TE. The tilted pillars of Fig. 9-b sketch this strain of the lattice, now storing *elastic energy* coming from the voltage V. This elastic energy would remain there while V and the cloud pulling it, *both exist*. This shows how a carrier would have converted into elastic energy of the lattice, the electrical energy $U_{Joul}$ loaded from the voltage V that it senses between terminals.

Since storing $U_{Joul}$ entails deforming a dipole of charge between the terminals of the resistor, the current entering its anode and leaving its cathode to pass from Fig. 9-a in TE to Fig. 9-b under *v(t)*=V, would be capacitive current that the generator sustaining V by the current $I_C$ should deliver to the $C_f$ of each carrier. Without such generator, (e. g. while measuring Johnson noise in TE), thermal activity would deliver randomly this current under the form of those fluctuations generating the voltage *v(t)* of Johnson noise. Sensing a bias voltage V in the resistor, each carrier would store the elastic energy $U_{Joul}$ ready to produce heat in $V_Q$. Although the small value of $C_f$ ($\approx 6\times10^{-18}$ Farad at room T) means that the energy loaded by each carrier is truly low, their high number in devices of macroscopic size leads to noticeable values of energy stored in this form.

An electric field E=V/*d* acting on $C_f$ with its charges -*q* and +*q* at mean distance S(T) kept by thermal activity, would increase this S(T) to superimpose the energy $U_{Joul}$ to its "ac energy" $U_f=q^2/(2C_f)$ of Eq. (9). Thus, the dipole of each carrier would store "dc energy" $U_{Joul}$ due to V and "ac energy" kT/2 J of thermal origin, in the same way the capacitor of a Sample & Hold system stores a small fluctuation of kT/2 joules (noise) around the voltage signal that it has sampled and that it holds for further processing. Thus, leaving aside the ac energy kT/2 J (noise) that sets $C_f$ in the absence of "signal" (i. e. for V=0), the "dc energy" loaded in $C_f$ by V (the signal) is given by Eq. (7).

Because electrons in resistors have several degrees of freedom (DOF) to take care of, the electronic cloud of a carrier will leave this DOF from time to time to access other DOF's, like becoming again the outer electron of a donor or becoming the electron that fills a surface state for example. In both cases we would say that "this carrier leaves the CB" to mean that its electronic cloud of charge disappears from the whole volume $V_Q$ to become highly located elsewhere. This allows speaking about $\tau_{CB}$ as the lifetime of an electron in the CB (i. e. its lifetime as a carrier). When a carrier loaded with $U_{Joul}$ leaves the CB (i. e. when it dies) the pulling action of its cloud ceases suddenly. The lattice thus released would start to vibrate as a spring-loaded lever suddenly released (Fig. 9-c) and the elastic energy $U_{joul}$ straining the lattice would pass to be a burst of vibrations injected to this periodic medium (phonons). This would end the process of converting electrical energy into heat propagating in the lattice of the conductive material.

This dissipation of $U_{Joul}$ each time the movable cloud of a carrier abandon $V_Q$ to become located elsewhere avoids any charge displacement that could modify the Johnson noise of the resistor. This reinforces the complementary roles of displacement and conduction currents in [2, 3]. Eq. (1) that was born to find the rate of fluctuations in a resistor of resistance R in TE [2] has led to this breaking proposal about the carrier of energy that dissipates its own energy each time it disappears. Eq. (8) states that loading energy in its carriers, the dissipation of a resistor will be increased without increasing its Johnson noise because dissipative currents are something else than currents generating noise (i. e. displacement ones). In this way the rate of carriers disappearing in a resistor is $\lambda$, both in TE or biased by a current $I_C$ if heating effects are negligible.

Using bias currents that take advantage of each carrier that disappears in a resistor to exist in it, we can explain why bias currents in resistors do not give noise, as it is found in experiments and as this work predicts. Arrived at this point, let me consider the type of fluctuation that would give a carrier if its movable cloud of charge appeared instantly located in terminal B. This would leave unscreened in $V_Q$ its PC of density $\rho^+=+q/V_Q$ C/cm$^3$ at first sight (see Fig. 10). The built-in voltage $\Delta\psi$ of this one-sided, space charge region supposedly left in $V_Q$, can be found from the field $E_{peak}=-q/(\varepsilon\times A_p)$ in the plate (terminal B) receiving the cloud of charge $-q$. It is the electric field suddenly found on the inner surface of terminal B of area $A_p$, varying linearly up to zero as we move towards terminal A. From the area of this triangle $\Delta\psi=(E_{peak}\times d)/2$ we have:

$$\Delta\Psi = V_A - V_B = -\frac{-q}{\varepsilon\times A_p}\times\frac{d}{2} = \frac{1}{2}\frac{q}{\frac{\varepsilon\times A_p}{d}} = \frac{1}{2}\frac{q}{C_{mat}} \quad \text{volts} \qquad (10)$$

Eq. (10) seems a solution of Poisson's equation to obtain the energy $E_1=q^2/(2C_{mat})$ that was used to obtain Eq. (1). However, the voltage $\Delta\psi=(V_A-V_B)$ that would appear suddenly between terminals in this case is not all the voltage shift $\Delta V=q/C_{mat}$ volts that we had used for fluctuations where a single electron was translocated between terminals. It only is $\Delta V/2$, thus warning us that "*an electron has arrived in terminal B without an electron having left simultaneously terminal A*." Put it bluntly: warning us that we are contravening the *current continuity* that the capacitive coupling of these two terminals at distance *d* does not allow. Thus, if the electronic cloud of a carrier becomes suddenly located in terminal A of a resistor, the electric field $E_{peak}=-q/(\varepsilon\times A_p)$ of its inner surface travelling towards terminal B at the speed of the electromagnetic wave, will set a field $E_{peak}=+q/(\varepsilon\times A_p)$ in its inner surface facing the former at distance *d*.

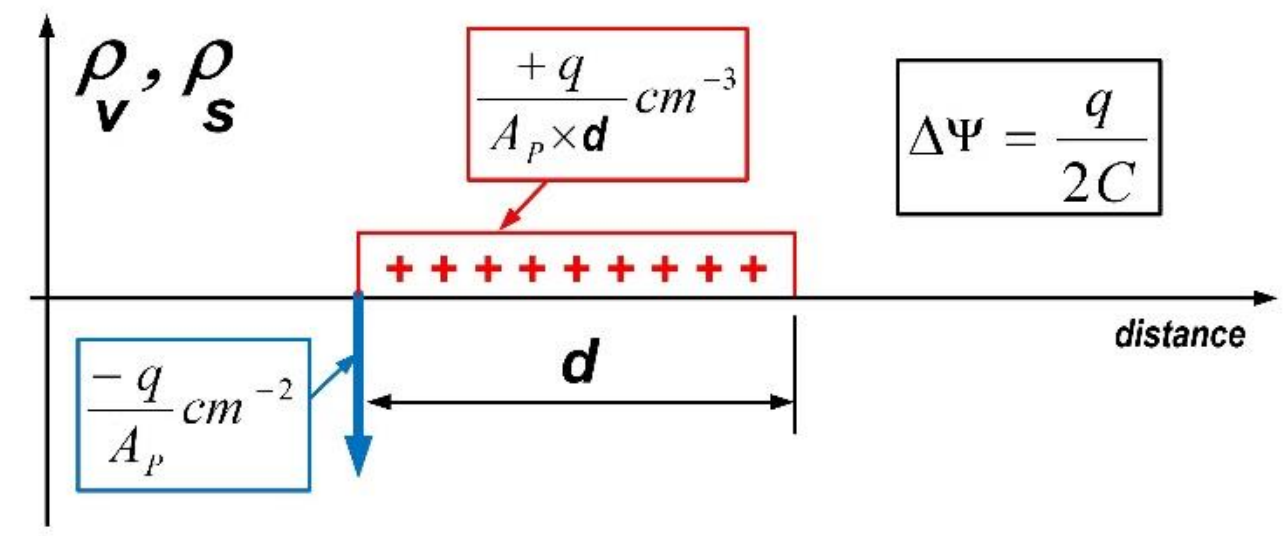

**Figure 10**. Charge densities if the electron cloud of a carrier could appear suddenly in terminal B of a resistor at finite distance *d* of its terminal A, leaving unscreened its PC.

Since these two fields of opposed sign appear *simultaneously* in different positions at distance *d* (at each terminal) due to their electromagnetic synchronization [3], I will consider that if the electronic cloud of a carrier disappears from $V_Q$ to appear located on one terminal as a surface density $-q/A_p$ C/cm$^2$, a density $+q/A_p$ C/cm$^2$ appears in the other terminal simultaneously. One would say that the unscreened PC of the carrier is "suddenly seen as a charge +*q*" on the inner surface of the other terminal. This doubles $\Delta\psi$ in Eq. (10) and solves the problem giving $\Delta\psi=q/C_{mat}$. This is an encouraging result suggesting that since we must use two terminals in electrical experiments, the PC of a carrier that I have proposed "ad hoc" intuitively, could be a form to imagine how the terminals of the resistor interact with the dipolar charge structure of a carrier.

Hence, if a carrier "dies" because its electronic cloud appears suddenly located in one terminal, it entails the simultaneous appearance of a charge +*q* in the other. Following our model, this instant translocation of an electron between terminals is the fluctuation [2-3] that, removing the negative cloud of a carrier between terminals, will release its energy to the lattice as heat. Given the huge number of majority carriers in macroscopic resistors, this fluctuation where a carrier ends its life as a carrier, would be dominant. About a carrier disappearing because its electronic cloud returns to a $Si^+$ cation or falls into a surface state, none of these capture events would give a fluctuation nor its shot noise, because no electron would arrive in one terminal. However, the energy of this carrier would be released as heat because any of these capture processes would remove its extended cloud of negative charge between terminals. Correspondingly, the emission of the electron thus trapped to the CB would not give shot noise.

As it is known, the lifetime $\tau_{CB}$ of electrons in the CB comes from different paths (e. g. trapping) that they can take to leave the CB. All of them contribute to define $\tau_{CB}$ whose evolution I am going to check briefly with devices that I have used. Despite its tiny value ($C_f=6{,}2\times10^{-18}$ F at room T) this capacitance per carrier can give high dissipation in macroscopic devices. Taking the cubic one of R=50Ω and $f_c$=32GHz of Section 3, its $V_Q=1\text{mm}^3$ would contain $n\approx10^{12}$ carriers ($n\approx N_d\times10^{-3}\text{cm}^3$ using full donor ionization at room T for this low $N_d$). From its R=50Ω and $C_{mat}$=0,1pF giving $\tau_d$=5ps, it would dissipate $P_{TE}=kT/\tau_d\approx828$pW in TE at room T. Biased by $I_C$=4mA, its voltage would be V=200mV, now dissipating $P_{DC}=V^2/R=800\mu$W ($P_{DC}\approx10^6$ times $P_{TE}$) because each carrier would load $U_{Joul}=1{,}24\times10^{-19}$ joules. Since the rate $\lambda=6{,}5\times10^{15}$ s$^{-1}$ of Eq. (1) for this resistor is the rate of carriers disappearing each second in this device with $n \approx 10^{12}$ carriers in all, their lifetime in the CB would be: $\tau_{CB}=n/\lambda=154\mu s$ on average.

Each carrier of this resistor living 154μs, would "die" 6500 times per second to dissipate $P_{TE}=828pW$ in TE or $P_{DC}=800\mu W$ when biased by $I_C=4mA$. Let me assume that 0,8mW do not heat noticeably this device of $V_Q=1mm^3$. Since $\tau_{CB}=154\mu s$ is much higher than $\tau_d=RC_{mat}=5ps$, each carrier of this resistor living 154μs on average, is plenty of time to load $U_{Joul}$ from V that only needs $t_{rise}=25ps$ ($5\tau_d$) to reach 99.3% of ($R\times I_C$) if $I_C$ was set suddenly. Connecting two cubic resistors in series gives a 100Ω resistor that for the same bias $I_C$, this longer device would dissipate twice the power of the cubic one of R=50Ω. Let me show how it agrees with our electronic noise. This double resistance means that the rate $\lambda^*$ of this longer resistor would be half the rate λ of the cubic resistor ($\lambda^*=\lambda/2$). Interestingly, the power that it would dissipate in TE at room T: $P_{TE}=kT/\tau_d=828pW$ is the same of the cubic one because keeping the material, $\tau_d$ and $P_{TE}$ both are kept.

The new voltage between terminals for $I_C=4mA$ is V=0,4 volts, thus the energy loaded in each carrier would be four times higher. Since its rate $\lambda^*$ of carriers that die is $\lambda^*=\lambda/2$, its dissipation $P_{dis}^*=\lambda^*\times 4U_{Joul}=2\times 800\mu W$ is twice that of the cubic resistor as expected. Because the number of carriers in this longer resistor has been doubled and their rate to disappear from the CB has been halved, their lifetime will be: $\tau_{CB}^*=n^*/\lambda^*$, thus $\tau_{CB}^*=4\tau_{CB}=616\mu s$. Interestingly, doubling the distance between terminals increases the carrier lifetime by four. After these results of our model "working in series" let me show its predictions for two cubic resistors of 50Ω connected in parallel. This gives a wider resistor of 25Ω whose power dissipated in TE would be: $P_{TE}=kT/\tau_d=828pW$ since its material and thus its $\tau_d$ is kept. However, this wider resistor of 25Ω biased by $I_C=4mA$, would dissipate half the power of the cubic resistor of 50Ω.

Due to its R/2=25Ω resistance, its rate λ* would be twice the rate λ of the cubic resistor (λ*=2λ). Because its voltage between terminals for $I_C=4mA$ only would be V=0,1 volt, the energy loaded in each carrier would be four times lower. Since the rate λ* of carriers that die in this resistor is: λ*=2λ, its dissipation becomes $P_{dis}^*=\lambda^*\times U_{Joul}/4=400\mu W$ as expected. Having doubled both the number of carriers in this wider resistor as well as their rate to disappear from the CB, their lifetime must be the same of the cubic resistor: $\tau_{CB}^*=n^*/\lambda^*=n/\lambda=154\mu s$. Interestingly, keeping the distance between terminals keeps the carrier lifetime. All in all, the idea of $Y(j\omega)$ collecting thermal energy by displacement currents between terminals to be dissipated by currents of its shunting conductance, would have a microscopic model.

I mean this one where electrons forming dipolar carriers in resistors convert electrical energy into elastic energy that is transferred to the lattice while the carrier exists. This energy is released as heat when the electronic cloud of the carrier leaves the CB, as it would occur each time a carrier disappears by a displacement current between its terminals. This is a sudden fluctuation of electric field that causes shot noise as well as a fluctuation of electrical energy held in the device. By contrast, if the cloud disappears to become located elsewhere within the volume of the device, its elastic energy is released as heat, but no fluctuation of electrical energy occurs. When the quantum of charge does not reach one terminal, it does not load electrical energy in its capacitance.